\documentclass[final,3p,times]{elsarticle}

\usepackage[utf8]{inputenc}
\usepackage{amsmath,amssymb,amscd,amsthm,amsfonts,eucal,dsfont,mathrsfs,latexsym}
\usepackage{graphicx} 
\usepackage{caption}
\usepackage{subcaption}
\usepackage{hyperref}

\hypersetup{
    bookmarks=true,         
    unicode=false,          
    pdftoolbar=true,        
    pdfmenubar=true,        
    pdffitwindow=false,     
    pdfstartview={FitH},    
    pdftitle={Estimation of Space Deformation Model for Non-stationary Random Functions},    
    pdfauthor={Francky Fouedjio},     
    pdfsubject={Non-stationary Random Functions},   
    pdfcreator={Francky Fouedjio},   
    pdfproducer={Francky Fouedjio}, 
    pdfkeywords={Non-stationary} {Variogram} {Deformation} {kernel smoothing} {kriging} {simulation} , 
    pagebackref=true,
    pdfnewwindow=true,      
    colorlinks=true,       
    linkcolor=red,          
    citecolor=green,        
    filecolor=magenta,      
    urlcolor=cyan          
}

\usepackage{sectsty}
\usepackage[compact]{titlesec} 
\allsectionsfont{\scshape\selectfont}
\usepackage{float}

\small\normalsize
\newtheorem{theorem}{Theorem}[section]

\newtheorem{prop}[theorem]{Proposition}

\theoremstyle{definition}

\theoremstyle{remark}

\journal{ }

\begin{document}

\begin{frontmatter}

\title{Estimation of Space Deformation Model for Non-stationary Random Functions}

\author[]{F. Fouedjio}
\ead{francky.fouedjio@mines-paristech.fr}

\author[]{N. Desassis}
\ead{nicolas.desassis@mines-paristech.fr}

\author[]{T. Romary}
\ead{thomas.romary@mines-paristech.fr}

\address{\'{E}quipe de G\'{e}ostatistique, Centre de G\'{e}osciences, MINES ParisTech}
\address{35, rue Saint-Honor\'{e}, 77305 Fontainebleau, France}

\begin{abstract}
Stationary Random Functions have been successfully applied in geostatistical applications for decades. In some instances, the assumption of a homogeneous spatial dependence structure across the entire  domain of interest is unrealistic. A practical approach for modelling and estimating non-stationary spatial dependence structure is considered. This consists in transforming a non-stationary Random Function into a stationary and isotropic one via a bijective continuous deformation of the index space. So far, this approach has been successfully applied in the context of data from several independent realizations of a Random Function. In this work, we propose an approach for non-stationary geostatistical modelling using space deformation in the context of a single realization with possibly irregularly spaced data. The estimation method is based on a non-stationary variogram  kernel estimator which serves as a dissimilarity measure between two locations in the geographical space. The proposed procedure combines aspects of kernel smoothing, weighted non-metric multi-dimensional scaling and thin-plate spline radial basis functions.
On a simulated data, the method is able to retrieve the true deformation. Performances are assessed on both synthetic and real datasets. It is shown in particular that our approach outperforms the stationary approach. Beyond the prediction, the proposed method can also serve as a tool for exploratory analysis of the non-stationarity.

\end{abstract}


\end{frontmatter}

\section{Introduction}
\label{sec1}

In the statistical analysis of spatial processes, modelling and estimating the spatial dependence structure is fundamental. It is used by prediction techniques like kriging or conditional simulations. Its description is commonly carried out using statistical tools such as the variogram or covariogram calculated on the entire domain of interest and considered under the stationarity assumption, for reasons of parsimony or mathematical convenience. The complexity of the spatial component of the analyzed process is therefore limited. 

The assumption that the spatial dependence structure is translation invariant over the whole domain of interest may be appropriate, when the latter is small in size, when there is not enough data to justify the use of a complex model, or simply because there is no other reasonable alternative. Although justified and leading to a reasonable analysis, this  assumption  is often inappropriate and unrealistic given certain spatial data collected in practice. Non-stationarity can occur due to many factors, including specific landscape and topographic features of the region of interest or other localized effects. These local influences can be observed computing local variograms, whose characteristics may vary across the domain of observations. For this type of non-stationarity structures, making spatial predictions using conventional stationary methods is not appropriate. Indeed, applying stationary approaches in such cases would be liable to produce less accurate predictions, including an incorrect assessment of the estimation error\citep{Ste99}.

Several approaches have been proposed to deal with non-stationarity through second order moments (see \citep{Gut13,Sam01,Gut94}, for a review). One of the most popular methods of introducing non-stationarity is the space deformation of \citet{Sam92} and other \citep{Mei97,Per98,Iov04}. It consists in starting with a stationary Random Function, and then transforming the distance in some smooth way to construct a non-stationary Random Function. Maximum likelihood and Bayesian variants of this approach have been developed by \citet{Mar93}, \citet{Smi96}, \citet{Dam01}, \citet{Sch03}. \citet{Per99,Per00,Per03}, \citet{Gen04}, \citet{Por10} established some theoretical properties about uniqueness and richness of this class of non-stationary models. Some adaptations have been proposed recently by \citet{Cas13}, \citet{Bor12}, \citet{Sch11}, \citet{Ver08,Ver09}.

A fundamental limitation of all estimation methodologies presented so far is the fact that implementation requires multiple independent realizations of the Random Function in order to obtain an estimated sample covariance or variogram matrix. The idea of having several independent realizations of the natural field is unrealistic because there are not multiple parallel physical worlds. In practice, the approach is feasible when a time series is collected at each location as this gives the necessary, albeit dependent, replications. In general, we would prefer to incorporate a temporal aspect in the modelling rather than attempting repairs (e.g., differencing and detrending) to achieve approximatively independent realizations.  However, many geostatistical applications involved only one measurement at each site or equivalently, only one realization of a Random Function.  \citet{And08}, \citet{And09} are the first authors to address the estimation of space deformation model in the case of a single realization of a Random Function, obtained as the transformation of a Gaussian and isotropic Random Function. They exhibit a methodology  based on quasi-conformal mappings and approximate likelihood estimation of the local parameters that characterize the deformation derived from partitioning densely observed data into subregions and assuming independence of the Random Function across partitions. However, this approach has not been applied to real datasets and requires very dense data.

In this work, we follow the pioneering work of \citet{Sam92}, while freeing the strong assumption of replication and do not make any distributional assumptions. In addition, we take into account other shortcomings associated with this approach that are: the required property of the deformation to be bijective and the computational challenge to fit the model for moderate and large datasets. To do so, we propose an estimation procedure based on the inclusion of spatial constraints and  the use of a set of representative points referred to as anchor points rather than all data points to find the deformation. The proposed method provides a non-parametric estimation of the deformation through a step by step approach: first a dissimilarity matrix is built by combining a non-parametric kernel variogram estimator and Euclidean distance between two points in the geographical space, second the estimation of the deformation at anchor points is considered by weighted non-metric multi-dimensional scaling and finally the deformation is interpolated over the whole domain by thin-plate spline radial basis functions. The proposed method also provides a rational and automatic estimation of the spatial dependence structure in the deformed space. We illustrate our estimation procedure on two simulated examples and apply it to soil dataset.

This paper is organized as follows: Section \ref{sec2} describes the space deformation model through its basic ingredients and main properties. In Sections \ref{sec3} and \ref{sec4}, we address the problem of estimating the non-stationary spatial dependence structure and  how spatial predictions and conditional simulations should proceed. Two simulated data in Section \ref{sec5} and a real dataset in Section \ref{sec6} are used to illustrate the performance of the new approach and its potential. Finally, Section \ref{sec7} outlines concluding remarks and further work.

\newpage 

\section{Space deformation representation}
\label{sec2}

The main idea behind the space deformation approach is the Euclidean embedding of a non-stationary Random Function into a new space of equal or greater dimension, where it can be easily described and modeled, that is to say where both stationarity and isotropy hold. 

Let $Z=\{Z(\mathbf{x}): \mathbf{x} \in G \subseteq \mathds{R}^p, p\geq 1\}$  be a constant mean Random Function  defined on a fixed continuous domain of interest $G$ of the Euclidean space $\mathds{R}^p$ and reflecting the underlying studied phenomenon. We consider that $Z$ is governed by the following model:
\begin{equation}{\label{Eq1}}
Z(\mathbf{x})=Y(f(\mathbf{x})), \ \forall \mathbf{x} \in G,
\end{equation}
which can be written equivalently:
\begin{equation}{\label{Eq2}}
Z(f^{-1}(\mathbf{u}))=Y(\mathbf{u}), \ \forall \mathbf{u} \in D,
\end{equation}
where $Y=\{Y(\mathbf{u}): \mathbf{u} \in D \subseteq \mathds{R}^{q},q\geq p\}$ represents a  stationary and isotropic Random Function and $f: G \rightarrow D$ represents a deterministic non-linear smooth bijective function of the $G$-space onto the $D$-space. In principle, we can allow $q\geq p$, although most frequently $q=p$. From now on, without loss of generality, we assume $q=p$.

The model specification in \eqref{Eq1} leads to model the variogram of $Z$ in the form:
\begin{equation}{\label{Eq3}}
\gamma(\mathbf{x},\mathbf{y})\equiv \frac{1}{2}\mathds{V}(Z(\mathbf{x})-Z(\mathbf{y}))=\frac{1}{2}\mathds{E}{(Z(\mathbf{x})-Z(\mathbf{y}))}^2=\gamma_0(\| f(\mathbf{x})-f(\mathbf{y}) \|),  \  \forall (\mathbf{x},\mathbf{y}) \in G \times G,
\end{equation}
where $\|.\|$ represents the Euclidean norm in $\mathds{R}^p$ and $\gamma_0(.)$ the  stationary   and isotropic variogram of $Y$ which depends only on the Euclidean distance between locations in $D$-space.

The second order structure model obtained in \eqref{Eq3} leads to a valid variogram, i.e. conditionally non-positive definite \citep{Mat71}. Its validity is assessed by the following proposition inspired by \citet{Mei97} and straightforward to establish (see \ref{appendix1}).

\begin{prop}\label{Prop1}
If $\gamma_0: D\times D \rightarrow \mathds{R}^+$ is a valid variogram, then $\gamma_0 \circ (f \times f)$ is a valid variogram on $G \times G$, for any function $f:G \rightarrow D$.
\end{prop}

It is then even possible to rely on a valid variogram in a different space, through a function linking the two spaces. Consequently, instead of working on the support $G$ of the non-stationary Random Function $Z$,  the variogram of $Z$ is defined with respect to the latent space $D$ where  stationarity and isotropy are assumed. Any problem involving the observed Random Function $Z$ is transposed by the deformation $f(.)$ to the stationary and isotropic Random Function $Y$. Standard geostatistical techniques, such as kriging and conditional simulations can be apply directly to the latter. The results obtained on $Y$ will then transpose to $Z$ by the inverse deformation $f^{-1}(.)$.

As described by \citet{Per99}, when $\gamma_0(.)$ is non-decreasing, the spatial deformation $f(.)$ operates as follows: the deformation effectively stretches the $G$-space in regions of relatively lower spatial correlation, while contracting it in regions of relatively higher spatial correlation, so that a stationary and isotropic variogram can model the spatial dependence structure as a function of the distance in the $D$-space representation.

It is also important to note that the spatial deformation model defined in \eqref{Eq3} is identifiable up to a scaling for $\gamma_0(.)$ and up to a homothetic Euclidean motion for $f(.)$. All isometries and homotheties are observationally equivalent. This result is based on the following proposition equivalent to that established by \citet{Per00} for deformation based non-stationary spatial correlation model. The proof is given in \ref{appendix1}.
 
\begin{prop}\label{Prop2}
If $(\gamma_0,f)$ is a solution to (\ref{Eq3}), then for any regular square matrix $\mathbf{A}$ and any vector $\mathbf{b}$, $(\tilde{\gamma_0},\tilde{f})$ with $\tilde{f}(\mathbf{x})=\mathbf{A}f(\mathbf{x})+\mathbf{b}$ and $\tilde{\gamma}_0(\|\mathbf{u}\|)=\gamma_0(\| \mathbf{A}^{-1}\mathbf{u}\|)$ is a solution as well.  
\end{prop}

\newpage
\section{Inference}
\label{sec3}
Let $\mathbf{Z}={(Z(\mathbf{s}_1),\ldots,Z(\mathbf{s}_n))}^T$ be a $(n \times 1)$ vector of observations from a unique realization of the Random Function $Z$, associated to known locations $\{\mathbf{s}_1,\ldots,\mathbf{s}_n\}\subset G\subseteq \mathds{R}^p$. In the second order structure model defined in (\ref{Eq3}), the functions $f(.)$ and $\gamma_0(.)$ are unknown and need therefore to be estimated. The estimation workflow involves four main steps. First, we define a non-stationary variogram non-parametric kernel estimator which serves as a dissimilarity measure between two points in the geographical space $G$. Second, we construct the deformed space $D$ using the procedure of weighted non-metric multi-dimensional scaling applied to a dissimilarity matrix built from the non-stationary variogram estimator. Third, we estimate $f(.)$ by interpolating between a configuration of points in the $G$-space and the estimations of their deformations in the $D$-space using the class of thin-plate spline radial basis functions. Fourth, the estimation of $\gamma_0(.)$  is carried out by calculating the experimental variogram on transformed data in the deformed space $D$ and using a mixture of basic variogram models, providing wide flexibility to adapt to the observed structure. 

\subsection{Non-stationary variogram estimator}
\label{ssec1}

We begin with an estimate of the non-stationary variogram \eqref{Eq3} at arbitrary locations using a kernel weighted local average of squared increments. Our proposal is to use the variogram cloud ${\gamma}_{ij}^{\star}=\frac{1}{2}{(Z(\mathbf{s}_{i})-Z(\mathbf{s}_{j}))}^2$ which is an unbiased estimator of $\gamma(\mathbf{s}_i,\mathbf{s}_j)$ \citep{Mat71} as the input data for the non-parametric kernel estimator of the non-stationary variogram. An intuitive empirical estimator of the non-stationary variogram at any two locations is given by the non-parametric kernel estimator defined as follows:
\begin{eqnarray}\label{Eq4}
\widehat{\gamma}(\mathbf{x},\mathbf{y};\lambda)& = & 
\frac{\sum_{i,j=1}^{n}{K_\lambda\left((\mathbf{x},\mathbf{y}),(\mathbf{s}_{i},\mathbf{s}_{j})\right)\left(Z(\mathbf{s}_{i})-Z(\mathbf{s}_{j})\right)^{2}}}{2\sum_{i,j=1}^{n}K_\lambda\left((\mathbf{x},\mathbf{y}),(\mathbf{s}_{i},\mathbf{s}_{j})\right)}{\mathds{1}}_{\{\mathbf{x} \neq \mathbf{y}\}}, \quad  \forall (\mathbf{x},\mathbf{y}) \in G \times G,
\end{eqnarray}

where $K_\lambda(.,.)$ is a kernel defined as: $K_\lambda\left((\mathbf{x},\mathbf{y}),(\mathbf{s}_{i},\mathbf{s}_{j})\right)=K(\mathbf{x},\mathbf{s}_{i};\lambda)K(\mathbf{y},\mathbf{s}_{j};\lambda)$, with $K(.,.;\lambda)$ a non-negative, symmetric kernel on $\mathds{R}^p \times \mathds{R}^p$ with bandwidth $\lambda \in \mathds{R}_+^*$. 

The expression \eqref{Eq4} defines the spatial dissimilarity between two arbitrary points in the geographical space $G$. The purpose of the kernel function is to weight observations with respect to a reference point so that nearby observations get more weight while remote sites receive less. The denominator of \eqref{Eq4} is a standardization factor that ensures $\widehat{\gamma}(\mathbf{x},\mathbf{y};\lambda)$ is unbiased when the expectation of the squared difference between the observations is spatially constant. If several realizations are available, we can simply take the average of \eqref{Eq4} over the different realizations.

Regarding the kernel $K(.,.;\lambda)$ defined in \eqref{Eq4}, his choice is less important than the choice of its bandwidth parameter $\lambda$, a reasonable choice of $K(.,.;\lambda)$ will generally lead to reasonable results \citep{Wan95}. We use  the quadratic kernel (Epanechnikov kernel) which is an isotropic kernel with compact support, showing optimality properties in density estimation \citep{Wan95}. Indeed, the computational cost of \eqref{Eq4} is greatly reduced by using a compactly supported kernel, as it reduces the number of terms to compute. The selection  of the bandwidth parameter $\lambda$ is adresssed in Section \ref{ssec4}. 

\subsection{Constructing the deformed space}
\label{ssec2}

The transformation of the geographical space $G$ into the deformed  space $D$ can be seen as the reallocation of a set of points in a Euclidean space (in a given dimensionality). The approach of non-metric multi-dimensional scaling (NMDS) provides a solution to this problem \citep{Kru64}. The aim is to find a representation of points in a given dimension space, such that the ordering of the Euclidean distances between points matches exactly the ordering of the dissimilarities.

\subsubsection{Anchor points}
\label{sssec1}

Till now, the fitting of a space deformation model based on NMDS procedure is a challenging numerical problem where the dimensionality is roughly proportional to the number of observations. A sample of size $n$ requires a $(n \times n)$ dissimilarity matrix to be stored and a $(n\times q)$ matrix of coordinates to be inferred. The search of the deformed space based on this dissimilarity matrix requires considerable computing time even when $n$ is moderately large. To reduce the computational burden, we can avoid mapping directly all data locations. Indeed, spatial dissimilarities calculated between pairs of close points can be unnecessary and redundant because of their high correlation. The idea consists in obtaining the deformed space using only a representative set of $m \ll n$ points referred to as anchor points over the geographical space $G$. These will be rather mapped with NMDS. Interpolating the anchor points in the $G$-space and the estimations of their deformations in the $D$-space, produces an estimate $\widehat f(.)$ of the deformation function. Then, the location of all sampling data points in the deformed space is obtained through $\widehat f(.)$. Representative points may be chosen as a sparse grid over the domain or a subset of data points. They allow to reduce the computation time and to give reliable results for the NMDS task. The sampling density which may be vary over the domain, can be accounted by non-uniform distribution of the anchor points.

\subsubsection{Dissimilarity matrix}
\label{sssec2}

We consider a set of $m$  $p$-dimensional anchor points $\mathbf{X}={[\mathbf{x}_{1},\ldots, \mathbf{x}_{m}]}^T$ of the geographical space $G$. The non-stationary variogram estimator $\widehat{\gamma}(.,.)$ defined in \eqref{Eq4} and calculated for each pair of anchor points allows to build a $(m \times m)$ symmetric dissimilarity matrix $\mathbf{\widehat{\Gamma}}_\lambda=[{\widehat{\gamma}_{ij}}(\lambda)]$, with ${\widehat{\gamma}_{ij}}(\lambda) = \widehat{\gamma}(\mathbf{x}_{i},\mathbf{x}_{j};\lambda)$. The specification of the dissimilarity matrix is required for the NMDS procedure. The latter is usually applied in the absence of the Euclidean coordinates of the points which produce the dissimilarities. However, in our context, the points of the original space $G$ are already identifiable by their Euclidean coordinates, hence the necessity to ensure the spatial consistency of the transformation. Moreover, $\widehat{\gamma}(.,.)$ reflects the local spatial dissimilarity, that is to say, for close pairs. Indeed, this estimator can be quite accurate for short and moderate distances, as in the stationary framework, but very imprecise, although well defined, for large distances. It is therefore necessary to penalize the importance given to large distances compared to short distances in the search of the deformation. Therefore, given the distance matrix associated with support points $\mathbf{X}$, we build a composite $(m \times m)$ dissimilarity matrix $\mathbf{\Delta}_{(\lambda,\omega)}=[\delta_{ij}(\lambda,\omega)]$:
\begin{equation}\label{Eq5}
\mathbf{\Delta}_{(\lambda,\omega)}=\omega\mathbf{\tilde{\Gamma}_\lambda} + (1-\omega)\mathbf{\tilde{D}}, \  \mbox{with } \mathbf{\tilde{\Gamma}_\lambda}=[\tilde{\gamma}_{ij}(\lambda)] \ \mbox{and } \mathbf{\tilde{D}}=[\tilde{d}_{ij}],
\end{equation}
where $\tilde{\gamma}_{ij}(\lambda)$ represents the scaled (transformed to the interval [0,1]) non-stationary variogram estimator at locations $\mathbf{x}_i$ and $\mathbf{x}_j$; $\tilde{d}_{ij}$ is the scaled Euclidean distance between $\mathbf{x}_i$ and $\mathbf{x}_j$; $\omega \in [0,1]$ is a mixing parameter.

The composite dissimilarity matrix $\mathbf{\Delta}_{(\lambda,\omega)}$ is a linear combination of a scaled dissimilarity matrix and a distance matrix. The idea is to build a hybrid spatial dissimilarity measure which takes into account both the dissimilarities observed in the regionalized variable and the spatial proximity to ensure that the deformation function does not fold, i.e. is bijective. Thus, the parameter $\omega$ controls the non-folding. The setting of $\omega$ is adressed in Section \ref{ssec4}.

\subsubsection{Fitting Non-Metric Multi-Dimensional Scaling Model }
\label{sssec3}

Given the $(m \times m)$ symmetric matrix $\mathbf{\Delta}_{(\lambda,\omega)}=[\delta_{ij}(\lambda,\omega)]$ of dissimilarities between the set of $m$  $p$-dimensional anchor points $\mathbf{X}={[\mathbf{x}_{1},\ldots, \mathbf{x}_{m}]}^T$, the objective is to represent $\mathbf{X}$ as a configuration of $q$-dimensional points $\mathbf{U}={[\mathbf{u}_{1},\ldots, \mathbf{u}_{m}]}^T$ such that the following relations are satisfied as much as possible:

\begin{equation}\label{Eq6}
\phi(\delta_{ij}(\lambda,\omega)) \approx \|\mathbf{u}_{i}-\mathbf{u}_{j}\| \equiv  h_{ij}(\mathbf{U}),
\end{equation}
where $\phi(.)$ is a monotonic function that preserves the rank order of the dissimilarities.

In other words, given the $\frac{1}{2}m(m-1)$ dissimilarities, we look for a representation of the anchor points in a given dimension space, so that the rank order of the configuration distances accord with the rank order of the dissimilarities. The representation of anchor points $\mathbf{U}$ in the deformed space $D$ is determined such that the distances between these points in the $D$-space minimize the loss function defined by a \textit{stress} measure:
\begin{equation}\label{Eq7}
S_{(\lambda,\omega)}(\mathbf{U})=\min_{\phi}{\left[\sum_{i < j}{\frac{{p_{ij}(\lambda)[\phi(\delta_{ij}(\lambda,\omega))-h_{ij}(\mathbf{U})]}^2}{\sum_{i < j}{p_{ij}(\lambda)h_{ij}^2(\mathbf{U})}}}\right]}^\frac{1}{2},
\end{equation}
where the minimization is performed over all monotonic functions $\phi(.)$ and $p_{ij}(\lambda)={\sum_{k,l=1}^{n}K_\lambda\left((\mathbf{x}_i,\mathbf{x}_j),(\mathbf{s}_{k},\mathbf{s}_{l})\right)}/{\|\mathbf{x}_i-\mathbf{x}_j\|}$,  for $i<j$. $K_\lambda(.,.)$ is the kernel function used in the computing of the non-stationary variogram estimator $\widehat{\gamma}(.,.;\lambda)$ defined in \eqref{Eq4}.
 
The estimator $\{{\phi}(\delta_{ij}(\lambda,\omega))\}$  represents the weighted least squares monotonic regression of the $\{h_{ij}(\mathbf{U})\}$ on the $\{\delta_{ij}(\lambda,\omega)\}$. $\{p_{ij}(\lambda)\}$ are non-negative dissimilarity weights used to weight the contribution of the corresponding elements of the dissimilarity matrix $\mathbf{\Delta}_{(\lambda,\omega)}$ in computing and minimizing \textit{stress}. We introduce such a weighting scheme  to take into account the fact that local information available in the neighborhood of anchor points may fluctuate from one anchor point pair to another. Hence highly reliable dissimilarities have more impact in the loss function than unreliable ones.  Such weights also allow to emphasize dissimalirities at short distances. 

The \textit{stress} provides a measure of the degree to which the rank order of the configuration distances concurs with the rank order of the dissimilarities to be mapped. It is invariant by translation, rotation or rescaling of the configuration. The problem defined in \eqref{Eq7} is solved by the Shepard-Kruskal NMDS iterative  algorithm as described, for example in, \citet{Bor05}, with the starting configuration taken as the configuration of anchor points $\mathbf{X}$.

\subsection{Parameter functions}
\label{ssec3}

In this paragraph, we are interested in the estimation of the functional parameters $f(.)$ and $\gamma_0(.)$ which govern the second order structure model \eqref{Eq3}. The transformation of anchor points is extended to a smooth function $\widehat{f}(.)$ over the entire geographical space $G$, using the class of thin-plate spline radial basis functions. Using the estimated deformation $\widehat{f}(.)$, an isotropic stationary variogram model is adjusted to the empirical variogram in the deformed space. Finally,  the estimated deformation function, in conjunction with the estimated isotropic stationary variogram model, are used to provide estimates of the spatial dependency between any two locations in the geographical space $G$.

\subsubsection{Deformation Mapping}
\label{sssec4}
We want to know the mapping of any point $\mathbf{x} \in G$ in the deformed space $D$, as well as  that in $G$ of any point $\mathbf{u} \in D$. Especially, a representation of all data points in the deformed space $D$ is required.

Given the configuration of anchor points $\mathbf{X}$ and the estimation of their deformation  $\mathbf{U}$, we seek a $q$-variate function $f:{\mathds{R}}^p \rightarrow {\mathds{R}}^q$ that satisfies the interpolation condition $f(\mathbf{X})= \mathbf{U}$. While a wide range of options exist, we follow \citet{Sam92} in using the class of thin-plate spline radial basis functions to estimate $f(.)$. Specifically, the thin-plate spline estimator of $f(.)$ takes the following form \citep{Dry98}:
\begin{equation}\label{Eq8}
\widehat f(\mathbf{x}) = (\widehat f_{1}(\mathbf{x}),\ldots,\widehat f_{q}(\mathbf{x}))^{T}
    = \mathbf{c} + \mathbf{A}\mathbf{x} + \mathbf{V}^{T}\boldsymbol{\sigma}(\mathbf{x}),
\end{equation}
where $ \mathbf{c}\mbox{ is }  (q\times 1), \mathbf{A}\mbox{ is }  (q\times p), \mathbf{V} \mbox{ is }  (m\times q), \boldsymbol{\sigma}(\mathbf{x})=(\sigma(\mathbf{x}-\mathbf{x}_1),\ldots,\sigma(\mathbf{x}-\mathbf{x}_m))^{T}$, $\mathbf{x}_1,\ldots,\mathbf{x}_m$ are the anchor points seen as the centers of the radial basis function
$\sigma(\mathbf{h})={\| \mathbf{h} \|}^2 \log(\| \mathbf{h} \|){\mathds{1}}_{\| \mathbf{h} \| > 0}$.

The parameters $\mathbf{c}, \mathbf{A}$ and $\mathbf{V}$ in the expression (\ref{Eq8})  are unknown and therefore have to be determined. This is achieved by resolving the system of equations $\widehat f(\mathbf{X})= \mathbf{U}$ under the constraints $\mathbf{1}^T\mathbf{V}=0 \mbox{ and } \mathbf{X}^T\mathbf{V}=0$ (see e.g. \citep{Dry98}).

\subsubsection{Transformed Variogram }
\label{sssec5}

The performance of the spatial deformation approach also depends on its ability to properly adjust an isotropic stationary variogram model in the deformed space. Thus, the modelling and estimating of the spatial dependence structure that characterizes the spatial behavior of the variable of interest is a key step. So far, highlighted by \citet{Mei97}, \citet{Per98}, \citet{Iov04}, no rational choice has  been developed yet for the isotropic stationary variogram model in the spatial deformation model. This choice is often made by visual inspection.

Here, the estimation of $\gamma_0(.)$ in the deformed space $D$ is carried out using a mixture of basic variogram models (nugget effect, gaussian, exponential, spherical,\ldots) thereby enriching the set of theoretical variograms and offering wide flexibility to match the observed structure:
\begin{equation}\label{Eq9}
\gamma_{0}(\| \mathbf{h} \|)=\sum_{k=1}^{K}c_k\gamma_k(\| \mathbf{h} \|), \quad \forall \mathbf{h} \in \mathds{R}^p,
\end{equation}
where $\gamma_1(.),\ldots,\gamma_K(.)$ are basic valid isotropic stationary variogram structures and $c_1,\ldots,c_K$ positive real numbers.

To estimate $\gamma_0(.)$, the isotropic stationary variogram model in the deformed space $D$, we use a  robust method recently developed by \citet{Des12}. It consists in automatically finding a model that fits the experimental variogram. From a linear combination of some authorized basic structures a numerical algorithm is used to estimate a parsimonious model that minimizes a weighted distance between the model and the experimental variogram.

\subsection{Tuning of hyper-parameters}
\label{ssec4}

The proposed approach relies on two hyper-parameters $(\lambda,\omega)$ used in the computation of the  composite dissimilarity matrix $\mathbf{\Delta}_{(\lambda,\omega)}$. The method used to select these hyper-parameters is data driven. We combine two forms of cross-validation to select their optimal value. 
 
The first form is the cross validation based on $\widehat{\gamma}(.,.)$. It consists of computing the cross-validation variogram error criterion over a grid of bandwidth values to determine a small set of bandwidths with low scores: 
\begin{equation}\label{Eq11}
CV_1(\lambda)=\frac{1}{n^2} \sum_{i,j=1}^n{\left(\widehat{\gamma}_{(-i,-j)}(\mathbf{s}_{i},\mathbf{s}_{j};\lambda)- {\gamma}_{ij}^{\star}\right)}^2,
\end{equation} 
where $\widehat{\gamma}_{(-i,-j)}(\mathbf{s}_{i},\mathbf{s}_{j};\lambda)$ is the non-stationary variogram estimator at locations $\mathbf{s}_{i}$ and $\mathbf{s}_{j}$ computed with all  observations except $\{Z(\mathbf{s}_{i}),Z(\mathbf{s}_{j})\}$. ${\gamma}_{ij}^{\star}=\frac{1}{2}{(Z(\mathbf{s}_{i})-Z(\mathbf{s}_{j}))}^2$ is an unbiased estimator of $\gamma(\mathbf{s}_i,\mathbf{s}_j)$.

Because the estimation of the spatial dependence structure is rarely a goal per se but an intermediate step before kriging, we want to choose the hyper-parameters that give the best cross-validation mean square error (MSE). More explicitly, for each pair of hyper-parameters $(\lambda,\omega)$, we compute the leave-one-out cross-validation score:
\begin{equation}\label{Eq12}
CV_2(\lambda,\omega)  =  \frac{1}{n}\sum_{i=1}^{n}{\left(Z(\mathbf{s}_{i}) - \widehat{Z}_{-i}(\mathbf{s}_{i};\lambda,\omega)\right)}^2,
\end{equation}
where $\widehat{Z}_{-i}(\mathbf{s}_{i};\lambda,\omega)$ denotes the spatial predictor computed at location $\mathbf{s}_{i}$ using all observations except $\{Z(\mathbf{s}_{i})\}$. The prediction method is described in detail in Section \ref{ssec5}.

The first form of cross-validation (\ref{Eq11}) allows to select a small grid of bandwidth values which are used in the second form of cross-validation (\ref{Eq12}) to search the optimal values of  $(\lambda,\omega)$. Thus, the first coarse bandwidth search reduces computational burden of the second step. This selection of hyper-parameters is illustrated with some examples in Sections \ref{sec5} and \ref{sec6}.

\section{Prediction}
\label{sec4}
The main purposes of modelling and estimating the spatial dependence structure is to spatially interpolate data  and perform simulations. We focus here on the description of the kriging and conditional simulations with the spatial deformation model \eqref{Eq3}.

\subsection{Kriging}
\label{ssec5}
The estimation of the space deformation model \eqref{Eq3} can be used in kriging-type estimation equations to provide more reliable spatial estimates and prediction standard errors than those based on an inadequate stationary variogram structure.

Given observations $Z(\mathbf{s}_1),\ldots,Z(\mathbf{s}_n)$ at $\mathbf{s}_1,\ldots,\mathbf{s}_n \in G$, the point predictor for the unknown value of $Z$ at unsampled location $\mathbf{s}_0 \in G$ is given by the ordinary kriging estimator:
\begin{equation}\label{Eq10}
\widehat{Z}(\mathbf{s}_{0})=\widehat{Y}(\mathbf{u}_{0})=\sum_{i=1}^{n}\alpha_{i}{(\mathbf{u}_{0})}Y(\mathbf{u}_{i}),
\end{equation}
which  minimizes the mean square error $\mathds{E}{(\widehat{Z}(\mathbf{s}_{0})-Z(\mathbf{s}_{0}))}^2$ under the constraint$:$ $\sum_{i=1}^{n}\alpha_{i}{(\mathbf{u}_{0})}=1$. $\mathbf{u}_{i}=\widehat{f}(\mathbf{s}_{i}), Y(\mathbf{u}_{i})=Z(\mathbf{s}_{i}), \ i=0,\ldots,n$ represents the transformed points.  

The prediction problem of the non-stationary Random Function $Z$ is then transposed to the isotropic stationary Random Function $Y$. Then, the kriging weights $(\alpha_{i}(\mathbf{u}_{0}), i=1,\ldots,n)$ are calculated by solving the well known ordinary kriging system \citep{Mat71} in the stationary framework. Thus, using the space deformation approach to estimate the regionalized variable on target grid points, we can proceed as follows:
\begin{enumerate}
\item obtain the image of the target grid points and data points through $\widehat {f}(.)$;
\item krige the transformed target grid points using $\widehat{\gamma}_0(.)$ and transformed data points;
\item obtain the kriging on the target grid points by simple correspondence.
\end{enumerate}

\subsection{Conditional simulations}
\label{ssec6}

There are a number of situations where conventional prediction techniques like kriging not properly fulfill their objectives. This is when it comes to integrating data, to predict quantities such as extrema or quantiles. An attractive approach to form consistent estimators of any entity of interest or quantifying uncertainties is through conditional simulations of the Random Function.

Here, the goal is to simulate a realization $\{z(\mathbf{s}), \mathbf{s} \in G \}$ of a Gaussian Random Function $Z$ with constant mean $m$ and second order structure model:
$\gamma(\mathbf{x},\mathbf{y})=\gamma_0(\| f(\mathbf{x})-f(\mathbf{y}) \|)$, with respect to the data $\{Z(\mathbf{s}_i)=z_i, i=1,\ldots,n\}$.

Perform conditional simulation of the non-stationary Gaussian Random Function $Z$ defined on the $G$-space means do the same for the Gaussian Random Function $Y$ defined on the $D$-space, with isotropic stationary variogram $\gamma_0(.)$. This can be accomplished based on a unconditional simulation $W$ of the Random Function $Y$ with same mean and variogram structure. For the latter, standard geostatistical simulation techniques already exist \citep{Lan02} (e.g. spectral method, turning band method ). Hence an algorithm of conditional simulation of the Random Function $Z$ is as follows:

\begin{enumerate}
\item Calculate the kriged estimates $y^{\star}(\mathbf{u})= m+ \sum_{i=1}^{n}\alpha_{i}{(\mathbf{u})}[y(\mathbf{u}_{i})-m]$ for each $\mathbf{u}=f(\mathbf{s}), \mathbf{s} \in G$;
\item Simulate  a Gaussian Random Function $W$ with constant mean $m$ and isotropic stationary variogram $\gamma_0(.)$ in the domain $D$ and at the transformed conditioning points. Let $\{w(\mathbf{u}), \mathbf{u}=f(\mathbf{s}),  \mathbf{s} \in G \}$ and $\{w_i=w(\mathbf{u}_i), \mathbf{u}_i=f(\mathbf{x}_i)\}$ be the generated values;
\item Calculate the kriged estimates $w^{\star}(\mathbf{u})= m+ \sum_{i=1}^{n}\alpha_{i}{(\mathbf{u})}[w_i-m]$ for each $\mathbf{u}=f(\mathbf{s}), \mathbf{s} \in G$;
\item Return $z(\mathbf{s})=y^{\star}(\mathbf{u})+ w(\mathbf{u}) - w^{\star}(\mathbf{u}), \mathbf{u}=f(\mathbf{s}),  \mathbf{s} \in G$.
\end{enumerate}

\section{Simulated examples}
\label{sec5}

To check whether our proposed estimation allows to find the correct deformation function, we apply it on one-dimensional example ($p=1$) and two-dimensional example ($p=2$).

\subsection{1D Example}
\label{ssec7}

We start with a one-dimensional simulation for which the results can be visualized easily. We generate at $1000$ irregularly spaced locations a zero mean, unit variance, Gaussian Random Function $Z$ over the domain $G=[0,1]$, based on the following second order structure model: $\gamma(x,y)=\gamma_0(|f(x)-f(y)|)$, where $\gamma_0(.)$ is an isotropic stationary exponential variogram model with range parameter $0.125$ and $f(x)=x^4$. Our goal is to estimate $f(.)$.

The deformation $f(.)$ leaves the domain $G$ globally unchanged but shrinks the left part of the segment whereas the right side is stretched. Thus, points of the segment which are located near the origin are highly correlated with their neighbors and points of the segment which are located near the extremity are slightly correlated with their neighbors. A set of $125$ regularly spaced anchor points are used to build the dissimilarity matrix. 

\begin{figure}[H]
        \centering
        \begin{subfigure}[h!]{0.27\textwidth}
                \centering
                \includegraphics[width=1\textwidth]{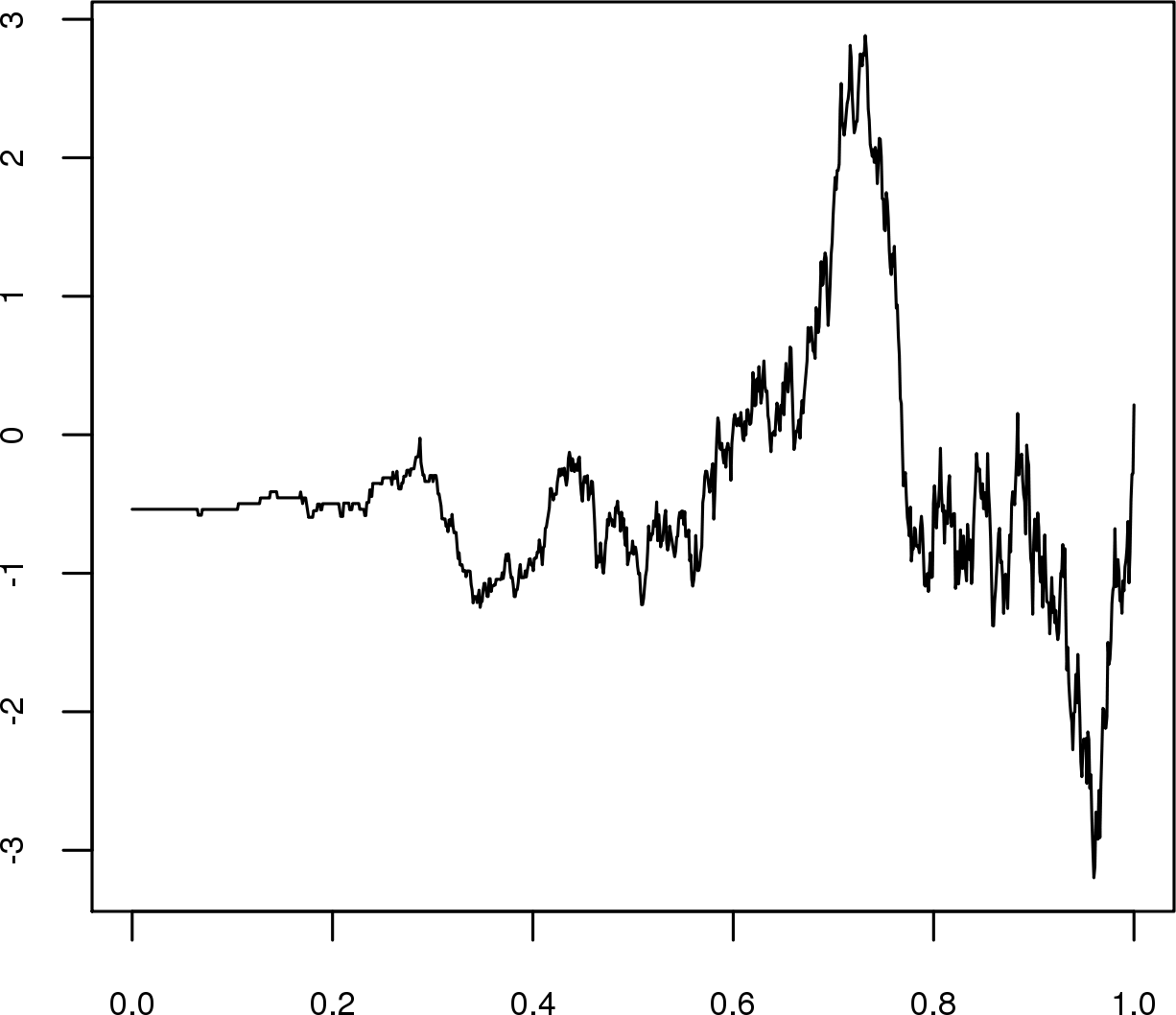}
                \caption{}\label{Fig1a}
        \end{subfigure}
         \quad     
        \begin{subfigure}[h!]{0.27\textwidth}
                \centering
                \includegraphics[width=1\textwidth]{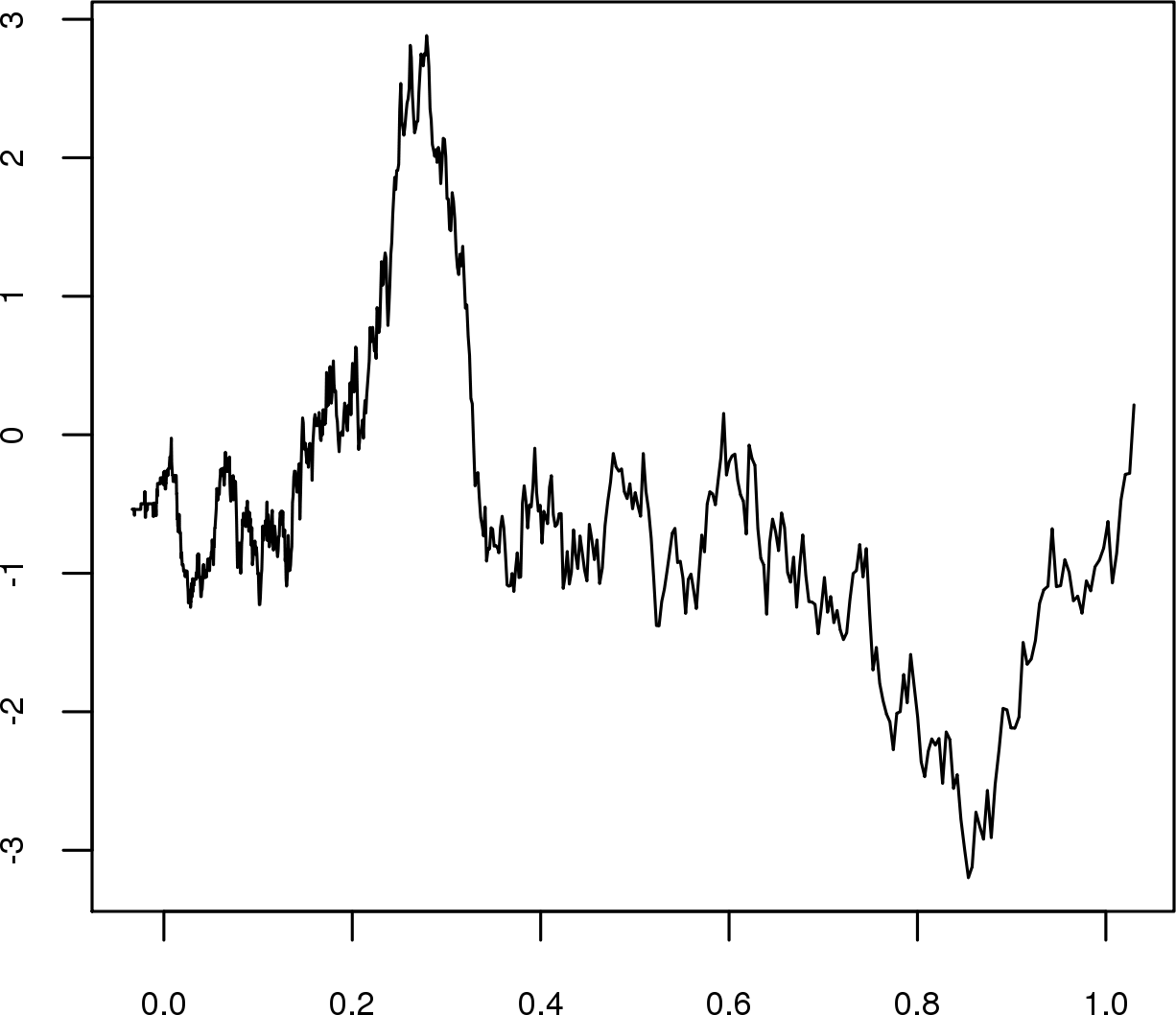}
                \caption{}\label{Fig1b}
        \end{subfigure}
        
        \medskip 
        \begin{subfigure}[h!]{0.27\textwidth}
                \centering
                \includegraphics[width=1\textwidth]{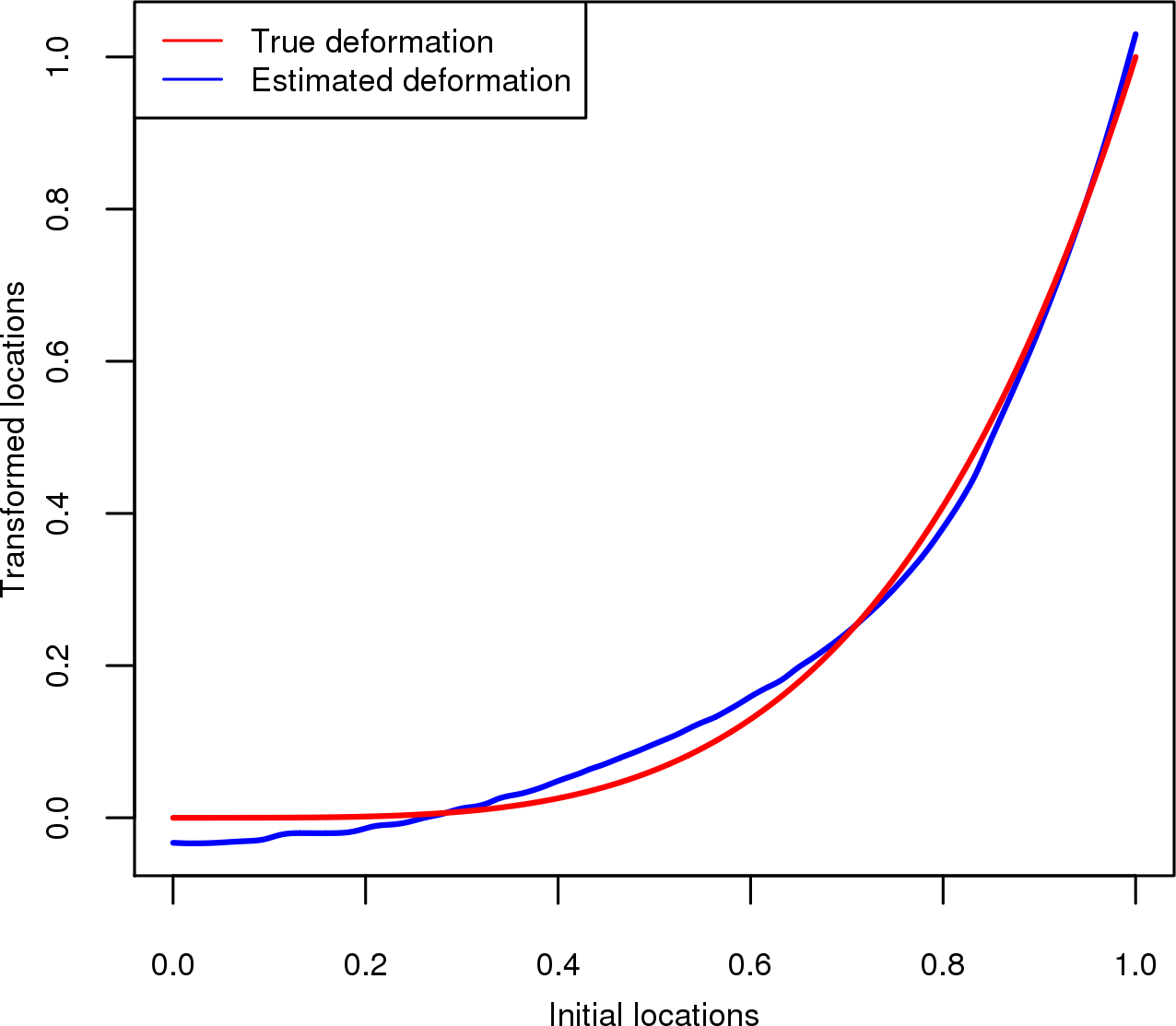}
                \caption{}\label{Fig1c}
        \end{subfigure}
        \quad 
        \begin{subfigure}[h!]{0.27\textwidth}
                \centering
                \includegraphics[width=1\textwidth]{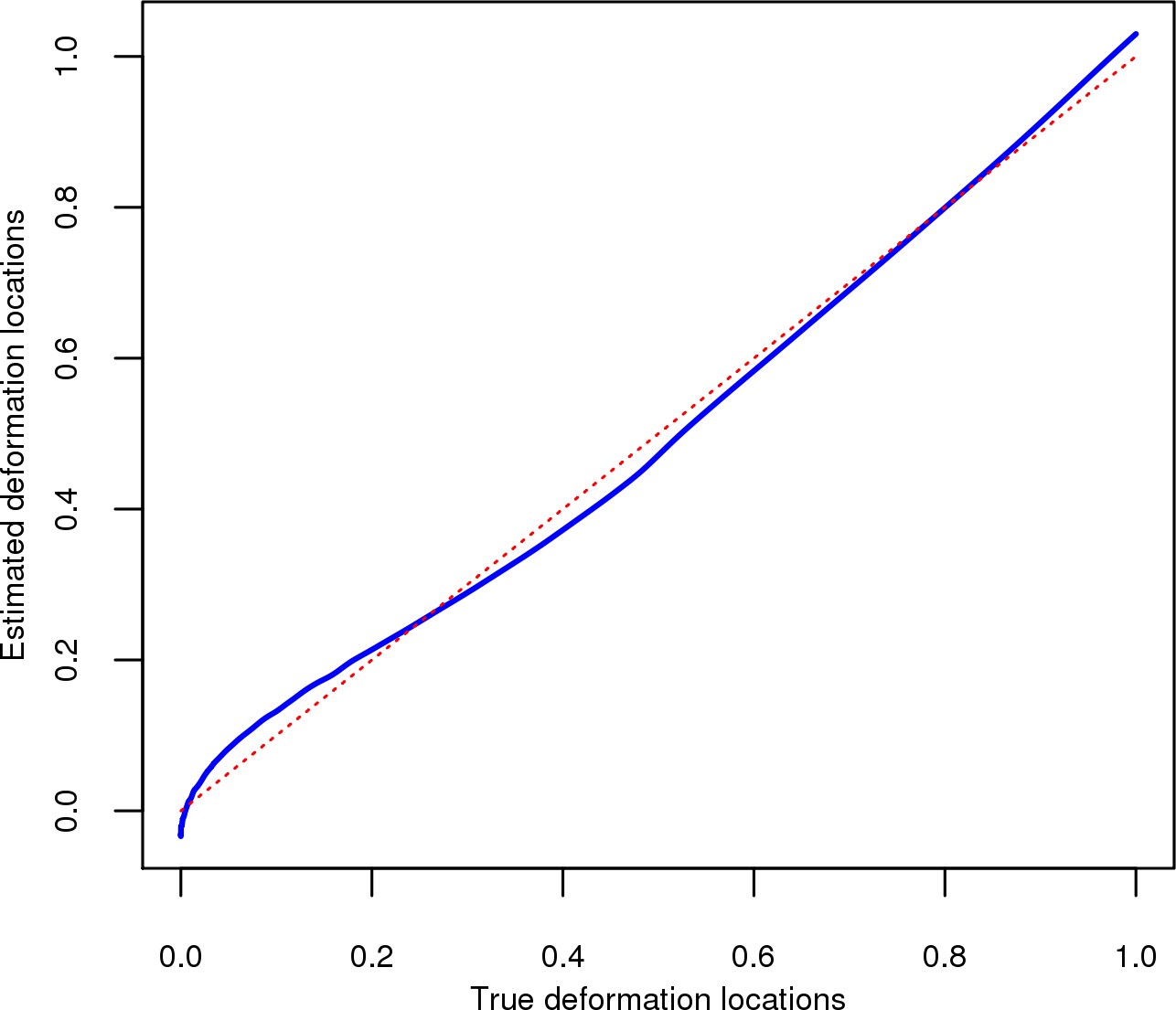}
                \caption{}\label{Fig1d}
        \end{subfigure}
        \caption{Example of a realization of the 1D non-stationary Random Function from space deformation model in the geographical space (a)  and estimated deformed space (b).  (c) True deformation and its estimation. (d) True deformation points versus estimated deformation points.}\label{Fig1}
\end{figure}

Figures \ref{Fig1a} and \ref{Fig1b} show a realization of the non-stationary Random Function $Z$ in the geographical space $G$ and in the estimated deformed space $D$.  The non-stationary process is transformed into stationary one simply by stretching and compressing the $x$-axe as shown in figure \ref{Fig1b}. The optimal value of hyper-parameters correspond to $\lambda=0.83$ and $\omega=0.65$. From the figures \ref{Fig1c} and \ref{Fig1d} we see that the estimated deformation is very close to the true deformation. This simple example illustrates the ability of the proposed approach to find the right deformation.

\subsection{2D Example}
\label{ssec8}
We consider a standardized Gaussian Random Function $Z$ over a domain $G={[0,1]}^2$, with a variogram: $\gamma(\mathbf{x},\mathbf{y})=\gamma_0(\| f(\mathbf{x})-f(\mathbf{y}) \|)\ $, where $ f(\mathbf{s})= \mathbf{o} + (\mathbf{s}-\mathbf{o})\| \mathbf{s}-\mathbf{o}\|$ is a radial basis function with center point $\mathbf{o}=(0.5,0.5)$ and $\gamma_0(.)$ an isotropic stationary cubic variogram model with range $0.05$. 

\begin{figure}[H]
        \centering
        \begin{subfigure}[h!]{0.28\textwidth}
                \centering
                \includegraphics[width=1\textwidth]{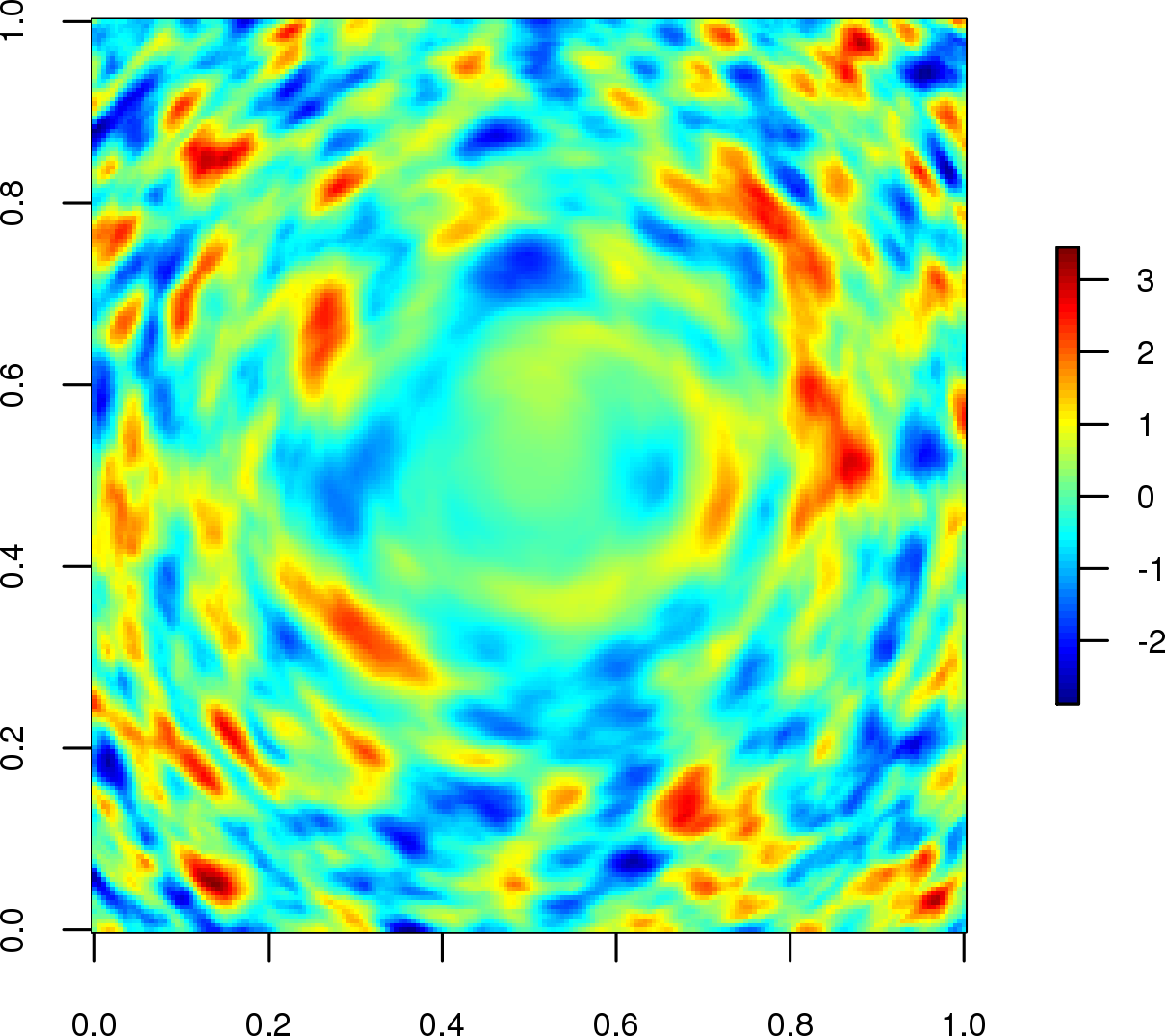}
                \caption{}\label{Fig2a}
        \end{subfigure}
        \quad 
        \begin{subfigure}[h!]{0.28\textwidth}
                \centering
                \includegraphics[width=1\textwidth]{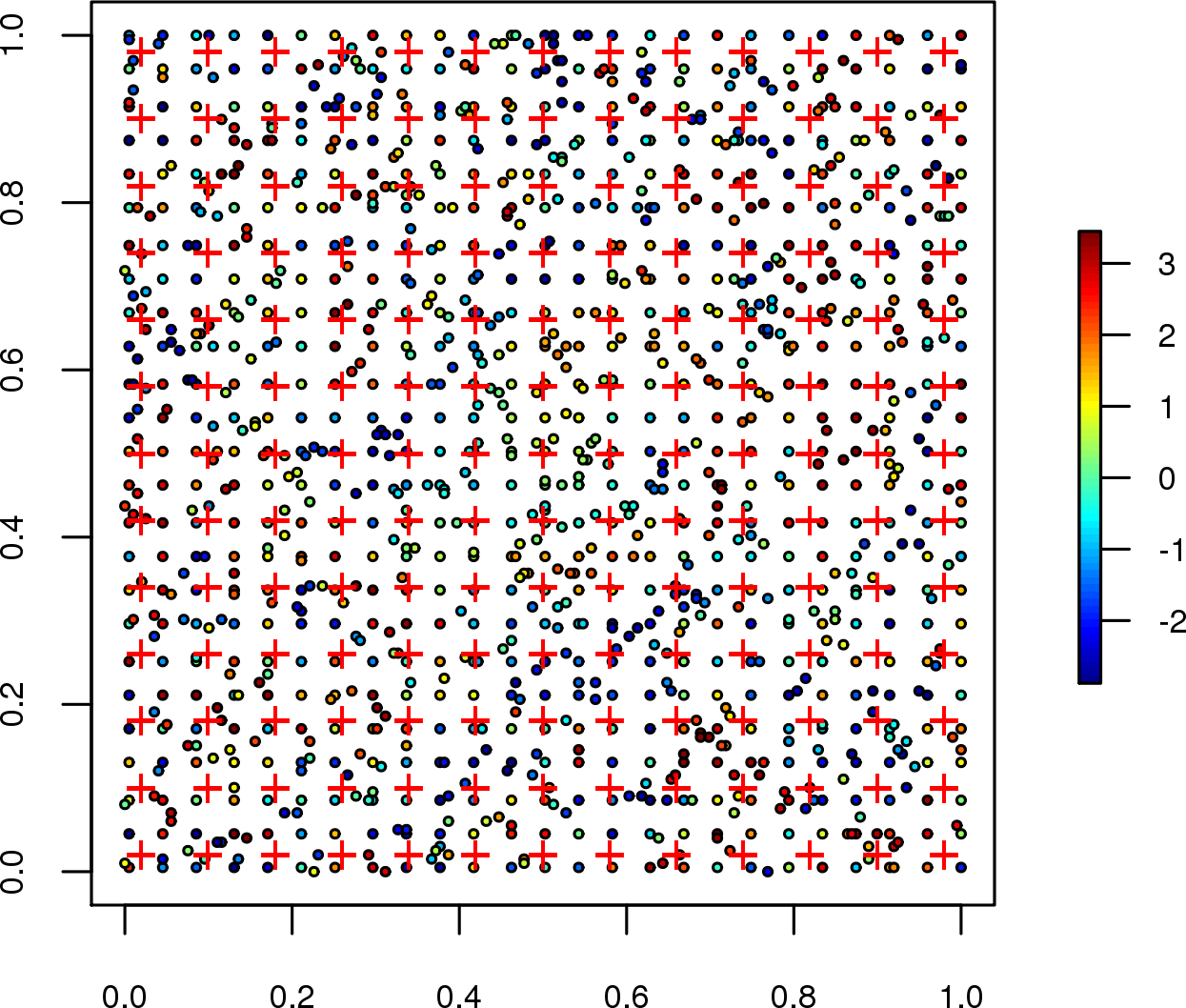}
                \caption{}\label{Fig2b}
        \end{subfigure}             
        
        \medskip
        \begin{subfigure}[h!]{0.28\textwidth}
                \centering
                \includegraphics[width=1\textwidth]{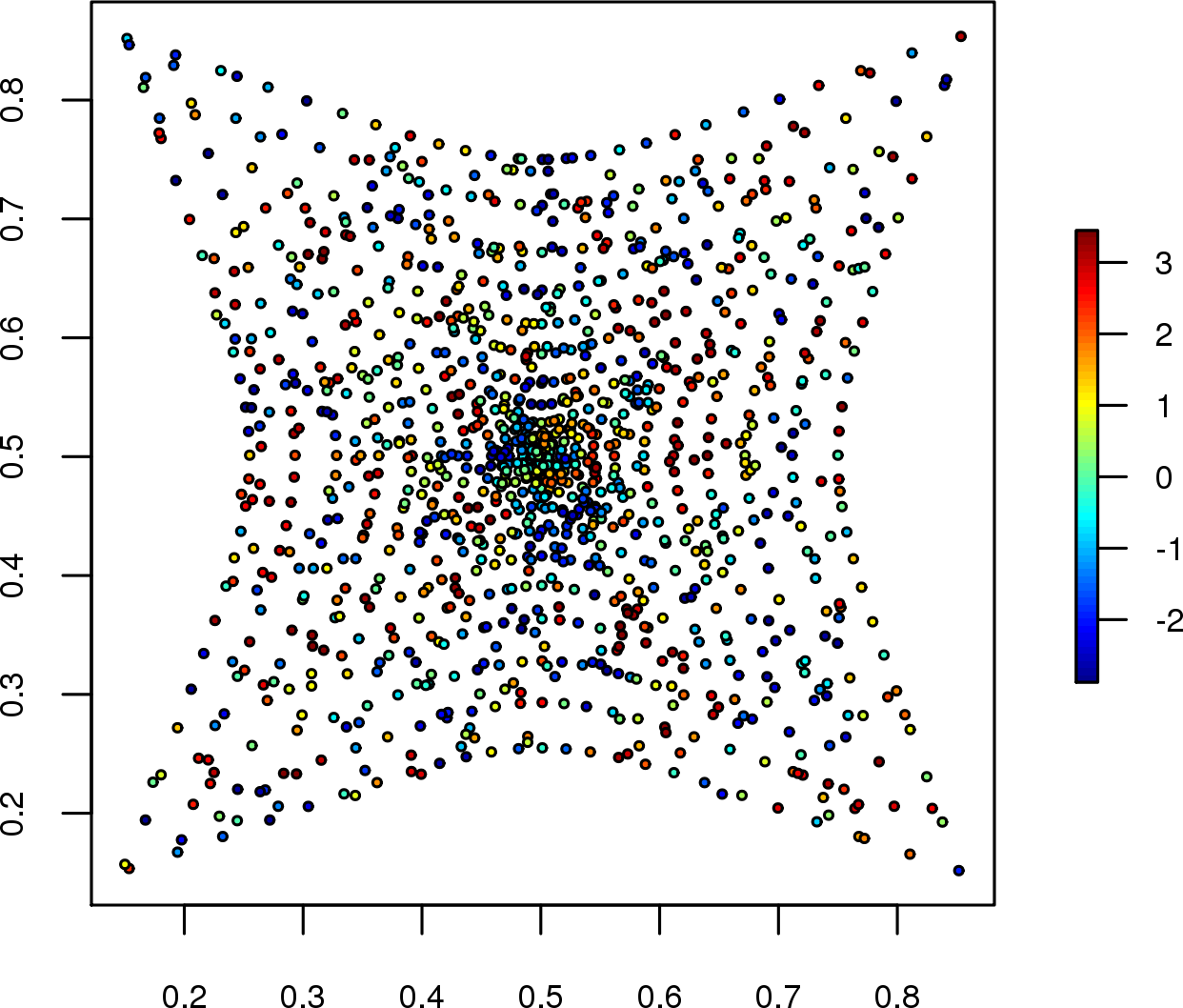}
                \caption{}\label{Fig2c}
        \end{subfigure}        
        \quad 
        \begin{subfigure}[h!]{0.28\textwidth}
                \centering
                \includegraphics[width=1\textwidth]{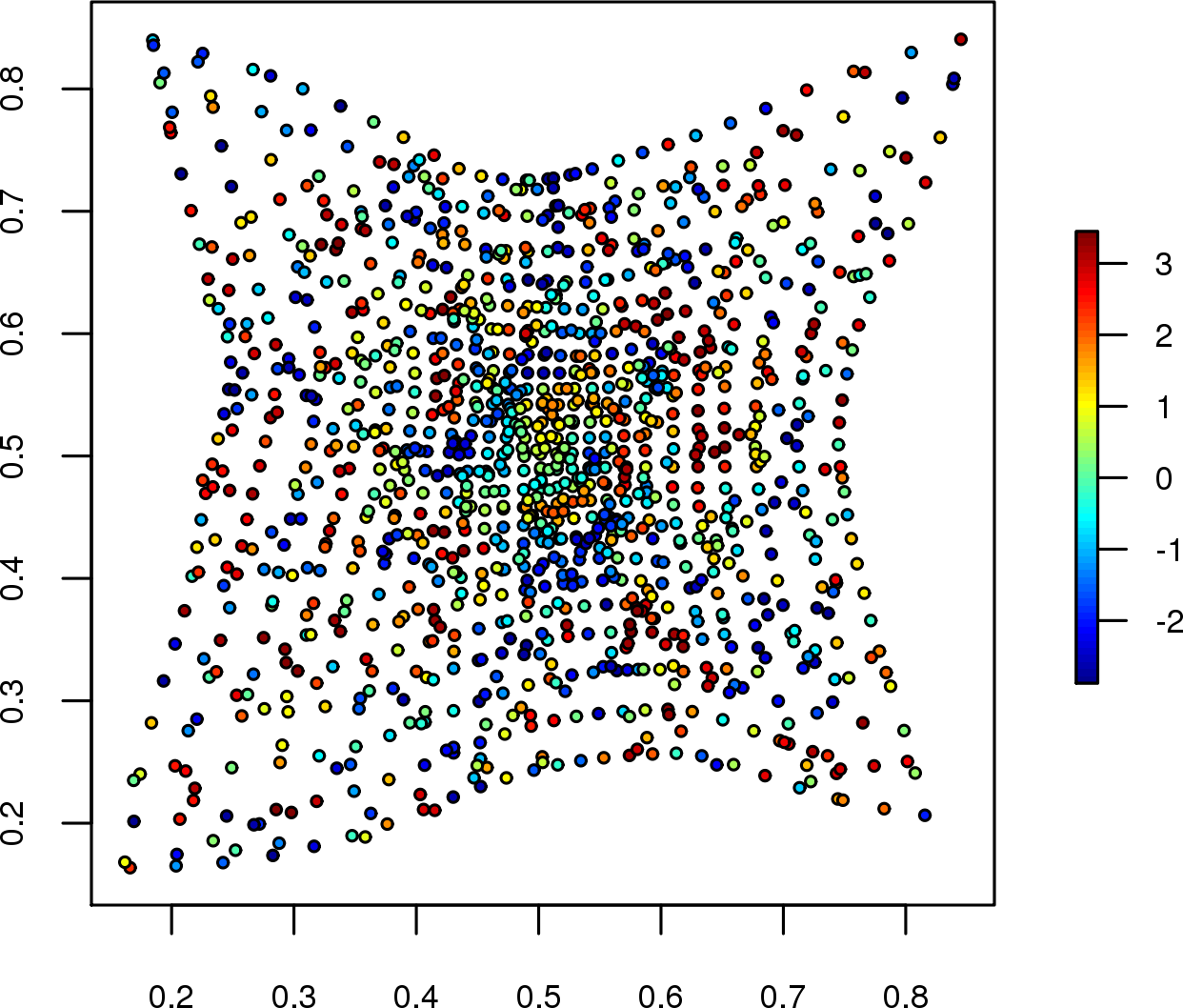}
                \caption{}\label{Fig2d}
        \end{subfigure}
        \caption{(a) Example of a realization of the 2D non-stationary Random Function from space deformation model in the geographical space. (b) sampling data and anchor points (red cross). (c, f) Sampling data points in the true deformed space and the estimated deformed space.}\label{Fig2}
\end{figure}

We simulate $Z$ at $200 \times 200$ points on a regular grid of $G$. Points of the domain which are located near the center are highly  correlated with their neighbors while points which are far from the center are slightly correlated with their neighbors (figure \ref{Fig2a}). $2249$ points are sampled and splitted into a $1225$ points training sample and a $1024$ points validation sample. We use a $(13\times13)$ regular grid as a configuration of anchor points (red cross points in figure \ref{Fig2b}).

Figure \ref{Fig2d} shows the estimated deformed space which looks similar as the true deformed space (figure \ref{Fig2c}). Our estimation method effectively stretches the domain in regions of relatively lower spatial correlation (at extremities) while contracting it in regions of relatively higher spatial correlation (at the center) so that an isotropic stationary variogram can model the spatial dependence structure in the deformed space. 

\begin{figure}[H]
        \centering
\begin{subfigure}[h!]{0.28\textwidth}
                \centering
                \includegraphics[width=1\textwidth]{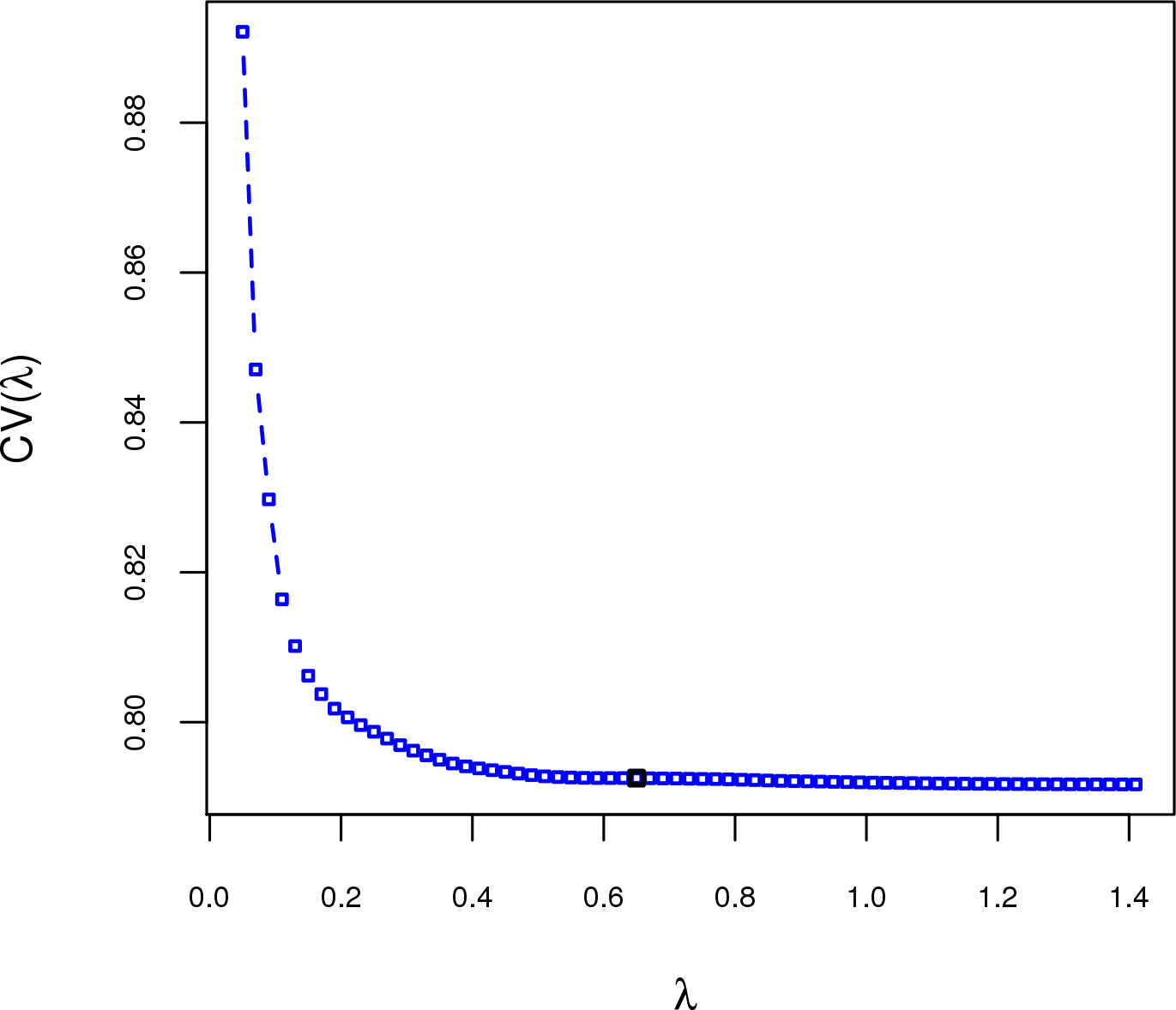}
                \caption{}\label{Fig3a}
        \end{subfigure}
        \qquad
        \begin{subfigure}[h!]{0.28\textwidth}
                \centering
                \includegraphics[width=1\textwidth]{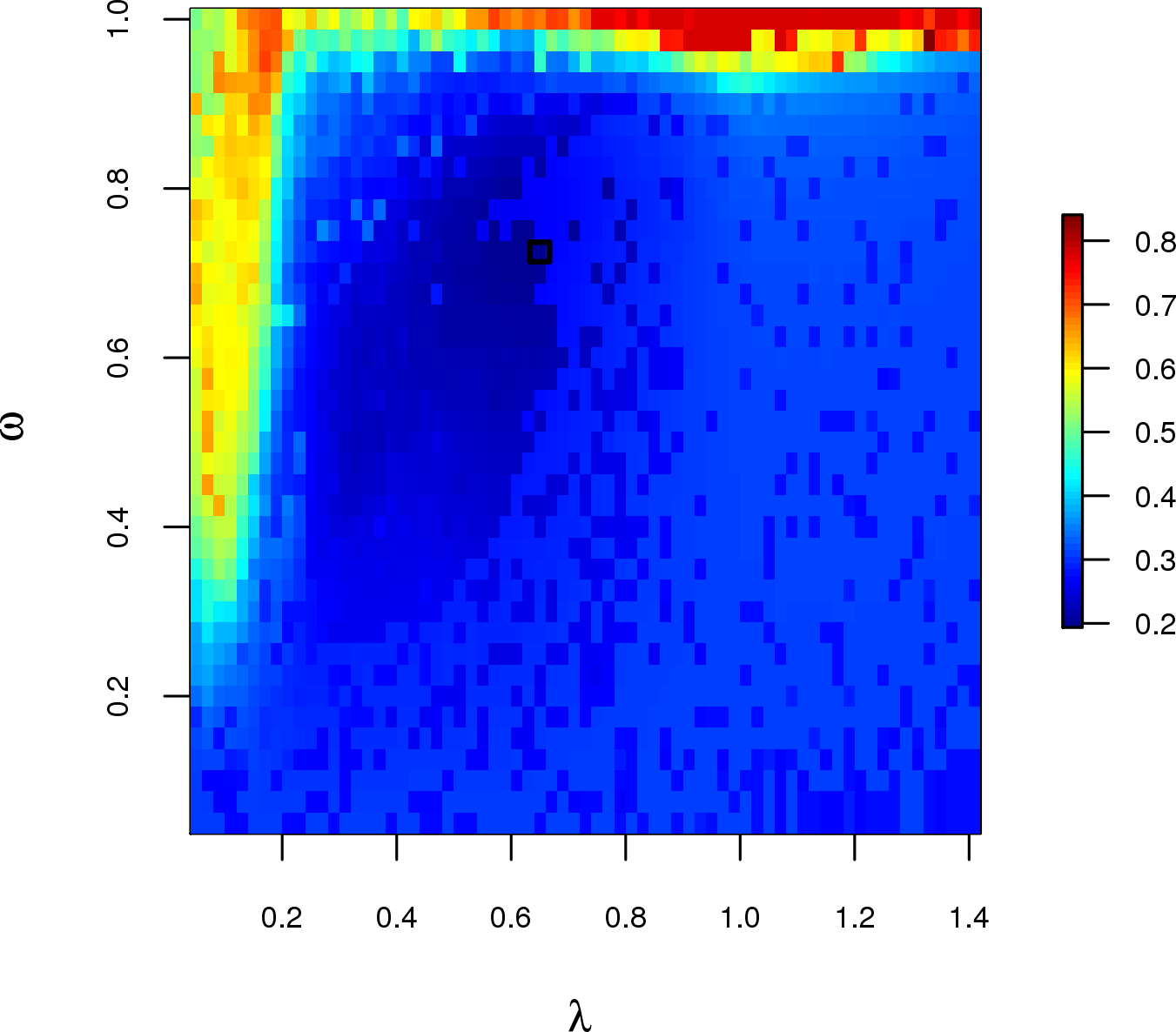}
                \caption{}\label{Fig3b}
        \end{subfigure}
        \caption{(a, b) Cross-validation score functions $CV_1(\lambda)$ and $CV_2(\lambda,\omega)$.}\label{Fig3}
\end{figure}

Figure \ref{Fig3a} gives the cross-validation variogram error function $CV_1(\lambda)$. We see that moderate and large values of $\lambda$ tend to give low  scores. Figure \ref{Fig3b} shows the second form cross-validation function $CV_2(\lambda,\omega)$. Its  minimization leads to $\lambda=0.65$ and $\omega=0.725$.

\begin{figure}[H]
        \centering
        \begin{subfigure}[h!]{0.25\textwidth}
                \centering
                \includegraphics[width=1\textwidth]{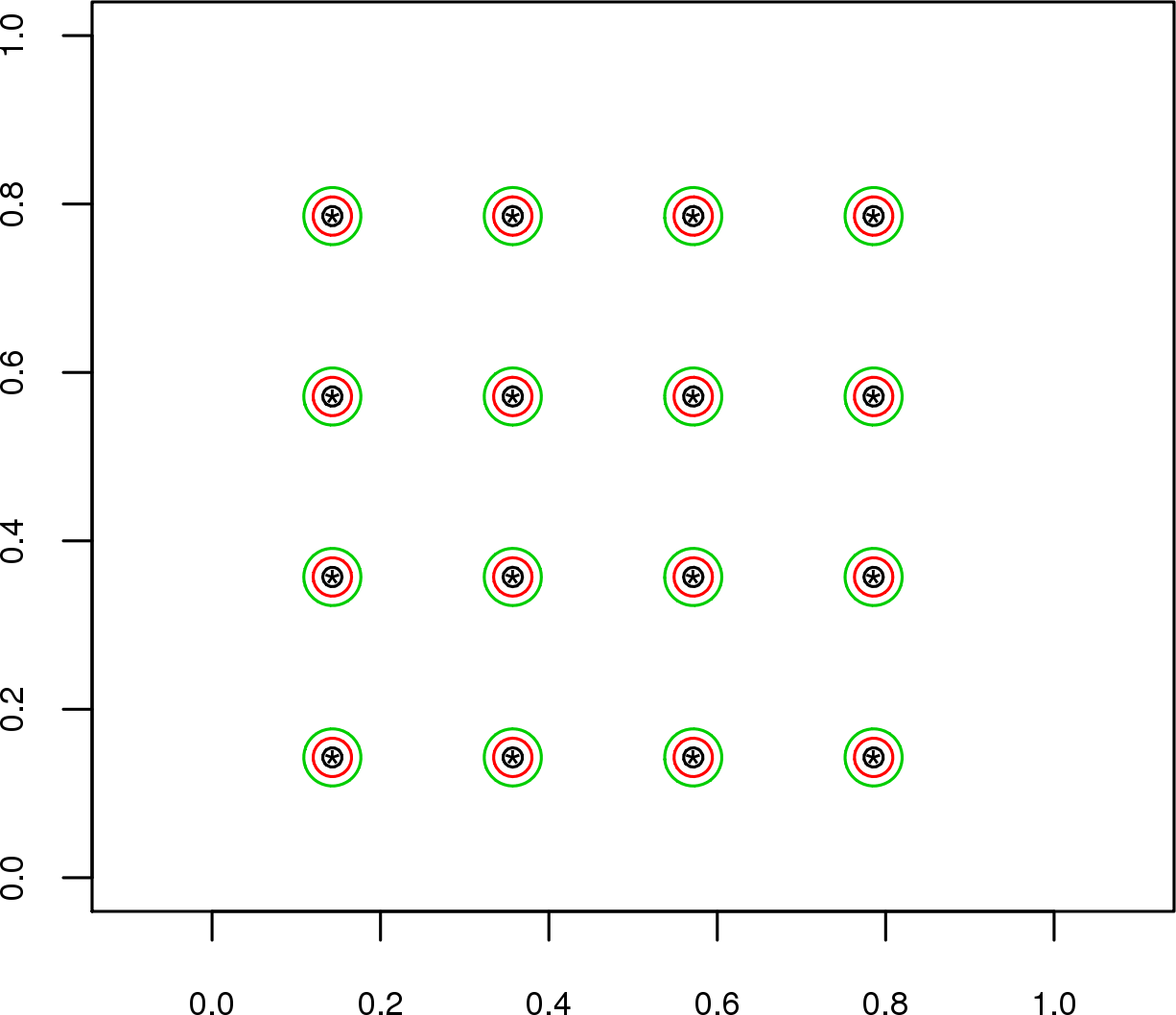}
                \caption{}\label{Fig4a}
        \end{subfigure}
        \quad
        \begin{subfigure}[h!]{0.25\textwidth}
                \centering
                {\includegraphics[width=1\textwidth]{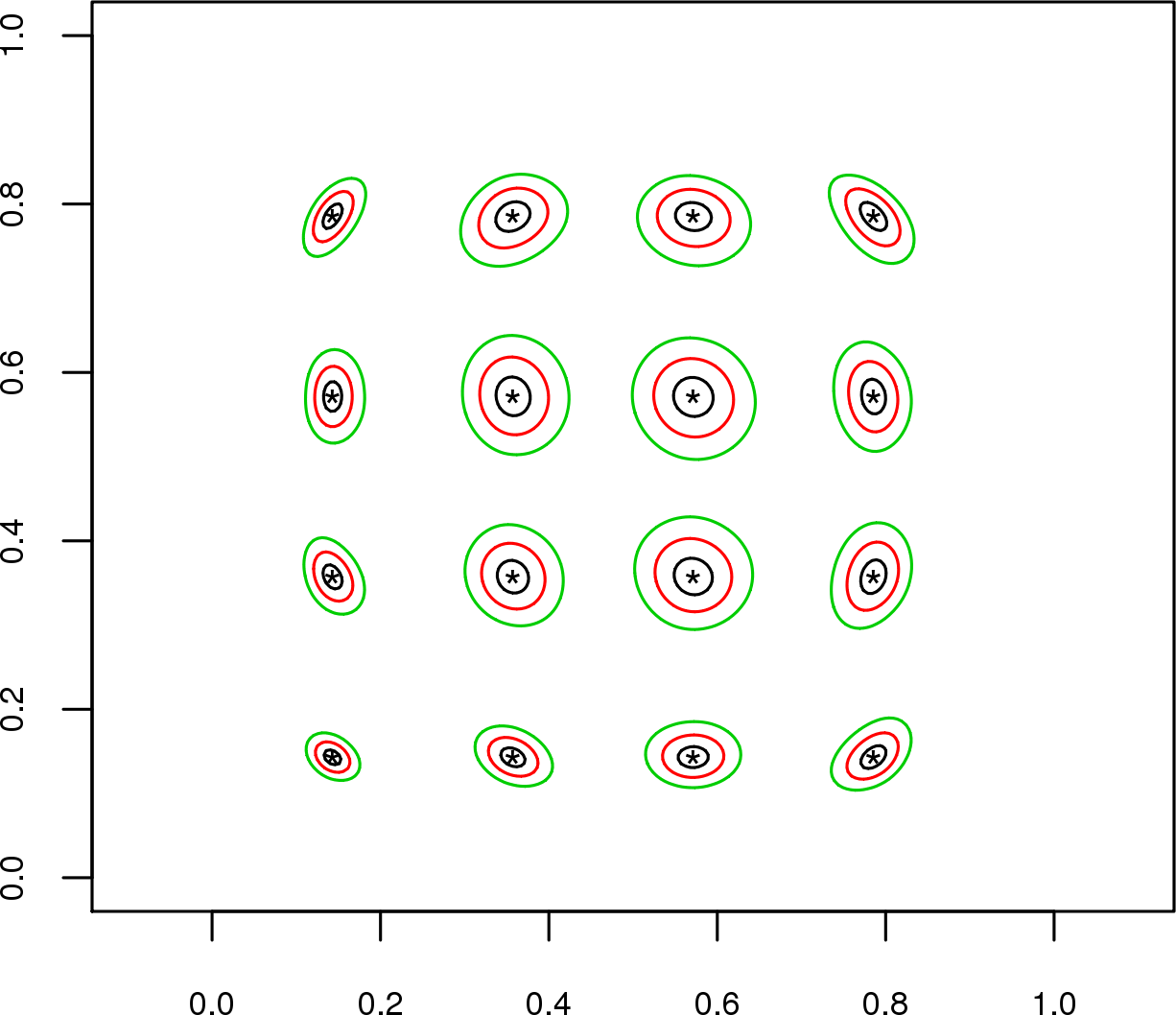}}
                \caption{}\label{Fig4b}
        \end{subfigure}
        \quad
        \begin{subfigure}[h!]{0.25\textwidth}
                \centering
                \includegraphics[width=1\textwidth]{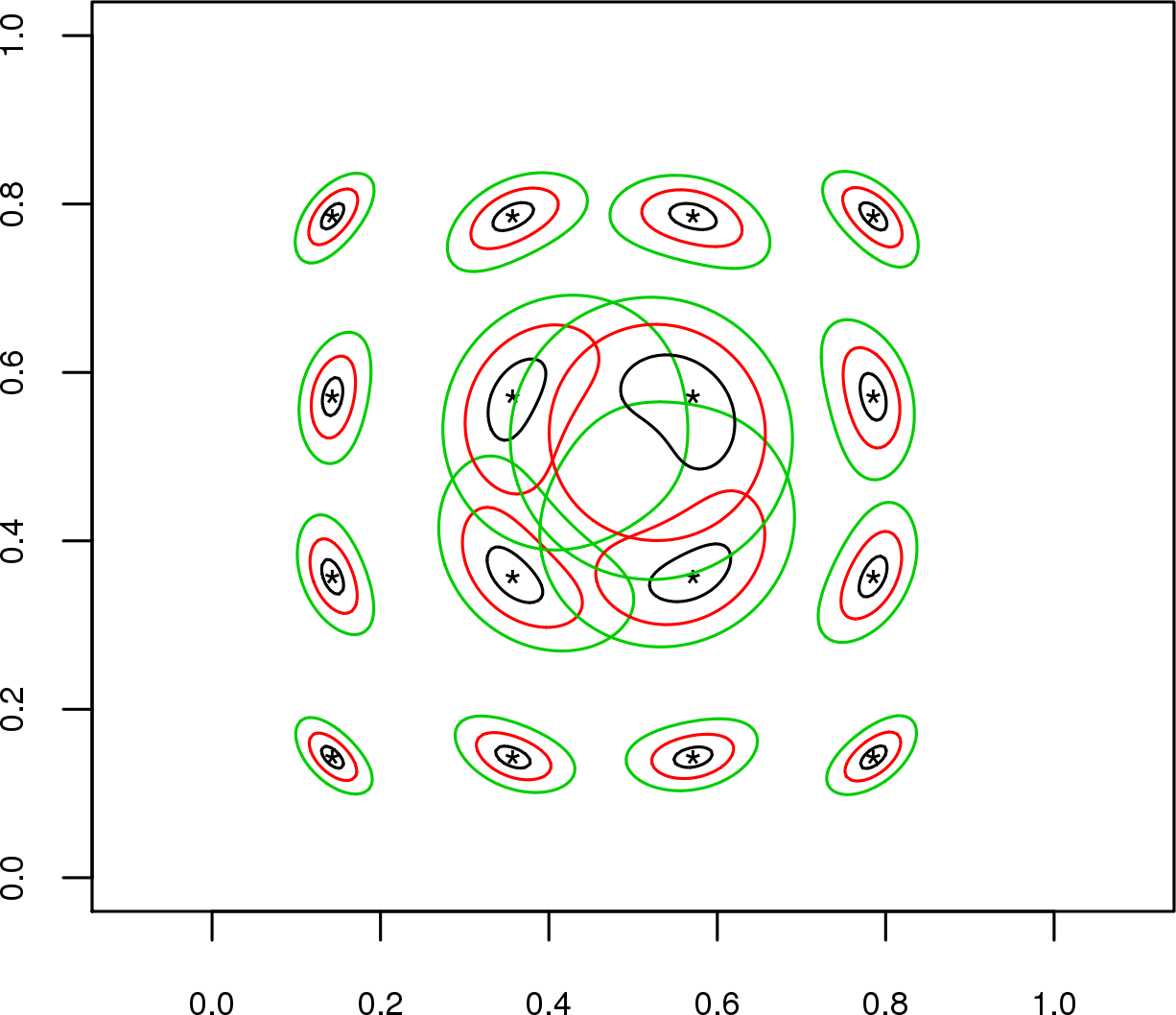}
                \caption{}\label{Fig4c}
        \end{subfigure}
        
        \caption{ Variogram  level contours at few points for the estimated stationary model, estimated non-stationary model and reference non-stationary model (a, b, c). Level contours correspond to the values: $0.1$ (black), $0.3$ (red) and $0.5$ (green).}\label{Fig4}
\end{figure}

A visualization of the variogram in a few points for  estimated stationary model, estimated non-stationary  model and non-stationary reference model is shown in figure \ref{Fig4}. From one site to another change in the non-stationary spatial dependence structure is clearly apparent.

To assess the predictive performance of our approach, the regionalized variable is predicted at the 1024 validation data points. Table \ref{Tab1} provides  a comparative performance of the estimated stationary model, the estimated non-stationary model and the reference non-stationary model. Some well-known discrepancy measures are used, namely the mean absolute error (MAE), the root mean square error (RMSE), the normalized mean square error (NMSE), the logarithmic score (LogS) and the continued probability score (CRPS). For RMSE, LogS and CRPS, the smaller the better; for MAE, the nearer to zero the better; for NMSE the nearer to one the better.

Table \ref{Tab1} summarizes the results for the predictive performance statistics computed on the validation data set. The stationary model is worse than the other two models for example, in terms of both RMSE, LogS and CRPS. The estimated non-stationary model performs comparably well as the reference model. The cost of not using a non-stationary model can be substantial. For example, the estimated stationary model is $18\%$ worse than the estimated non-stationary model in terms of RMSE.

\begin{table}[H]
\begin{center}
\begin{tabular}{lccc}
\hline
    & Stationary Model& Non-Stationary Model & Reference Model   \\
         \hline
    
    Mean Absolute Error   & 0.28   & 0.24 & 0.23\\
            
    Root Mean Square Error  & 0.44    & 0.37 & 0.35 \\
    
    Normalized Mean Square Error & 2.03  & 1.17 & 1.08 \\

    Logarithmic Score & 1118   & -69 & -92  \\
   
    Continued Rank Probability Score & 0.35   & 0.29 & 0.28   \\
    \hline
\end{tabular}
\end{center}
\caption{Predictive performance statistics on a test set of 1024  locations.}
\label{Tab1}
\end{table}

\begin{figure}[H]
        \centering
        \begin{subfigure}[h!]{0.28\textwidth}
                \centering
                \includegraphics[width=1\textwidth]{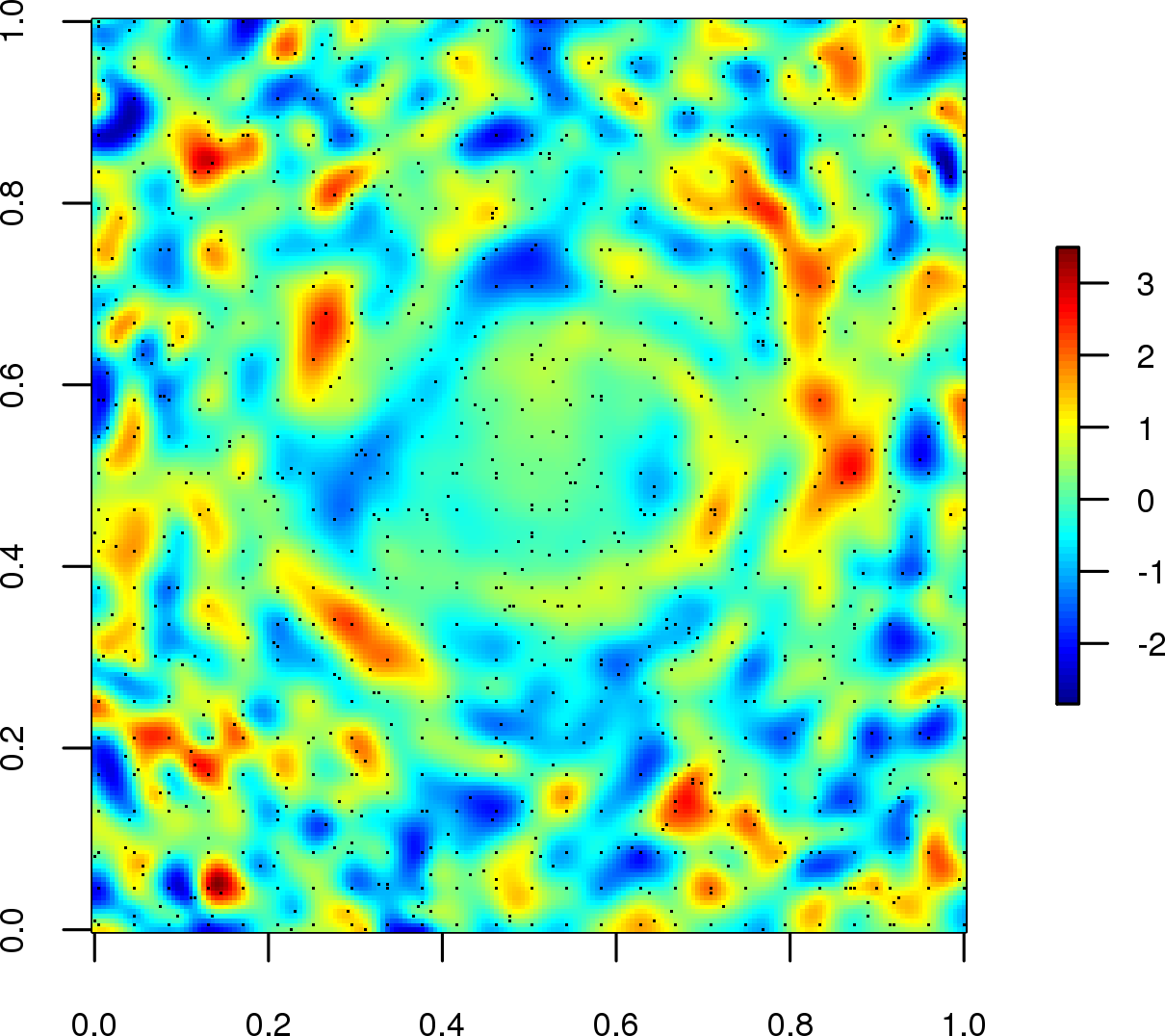}
                \caption{}\label{Fig5a}
        \end{subfigure}
        \quad
        \begin{subfigure}[h!]{0.28\textwidth}
                \centering
                {\includegraphics[width=1\textwidth]{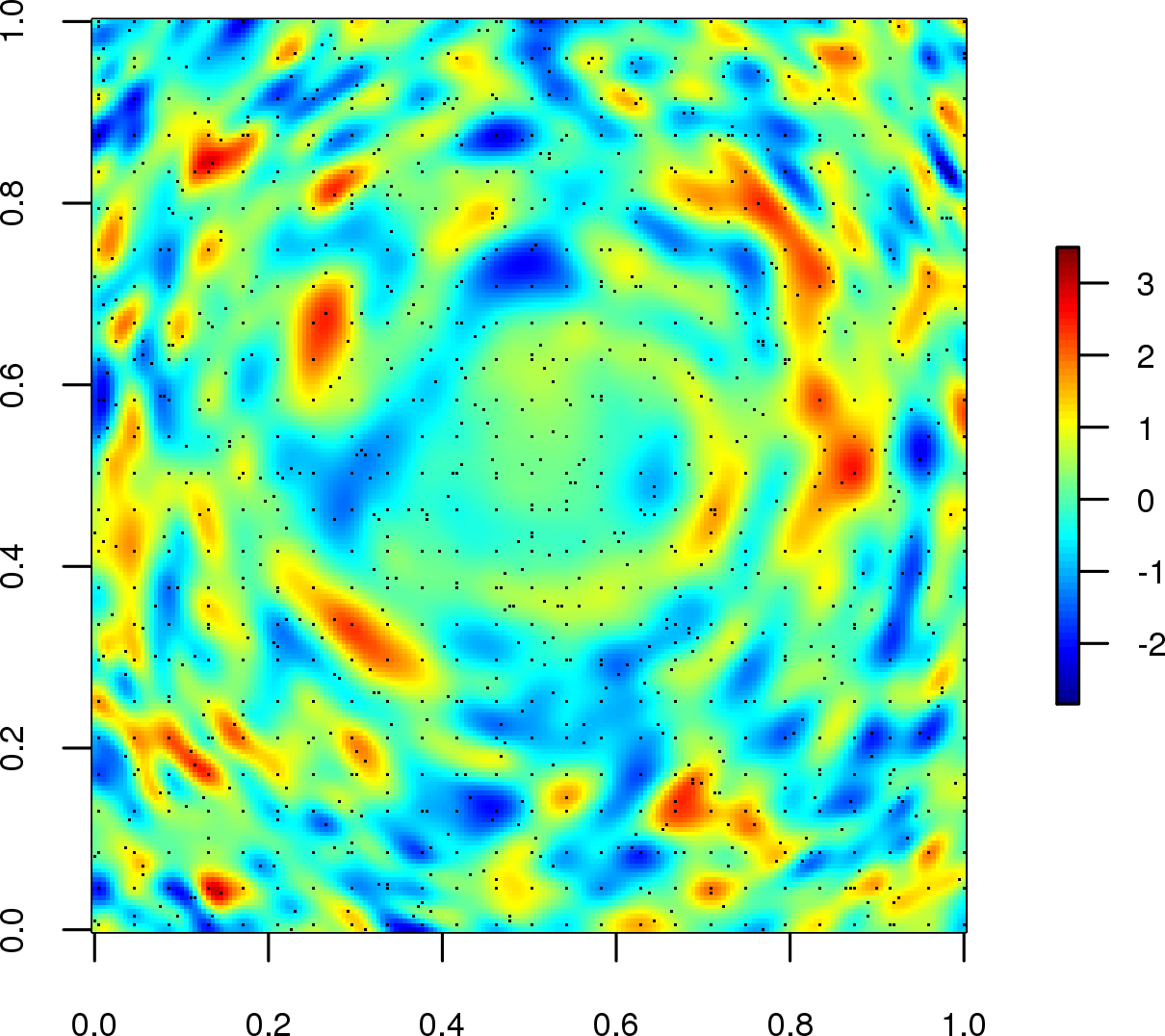}}
                \caption{}\label{Fig5b}
        \end{subfigure}
        \quad
        \begin{subfigure}[h!]{0.28\textwidth}
                \centering
                \includegraphics[width=1\textwidth]{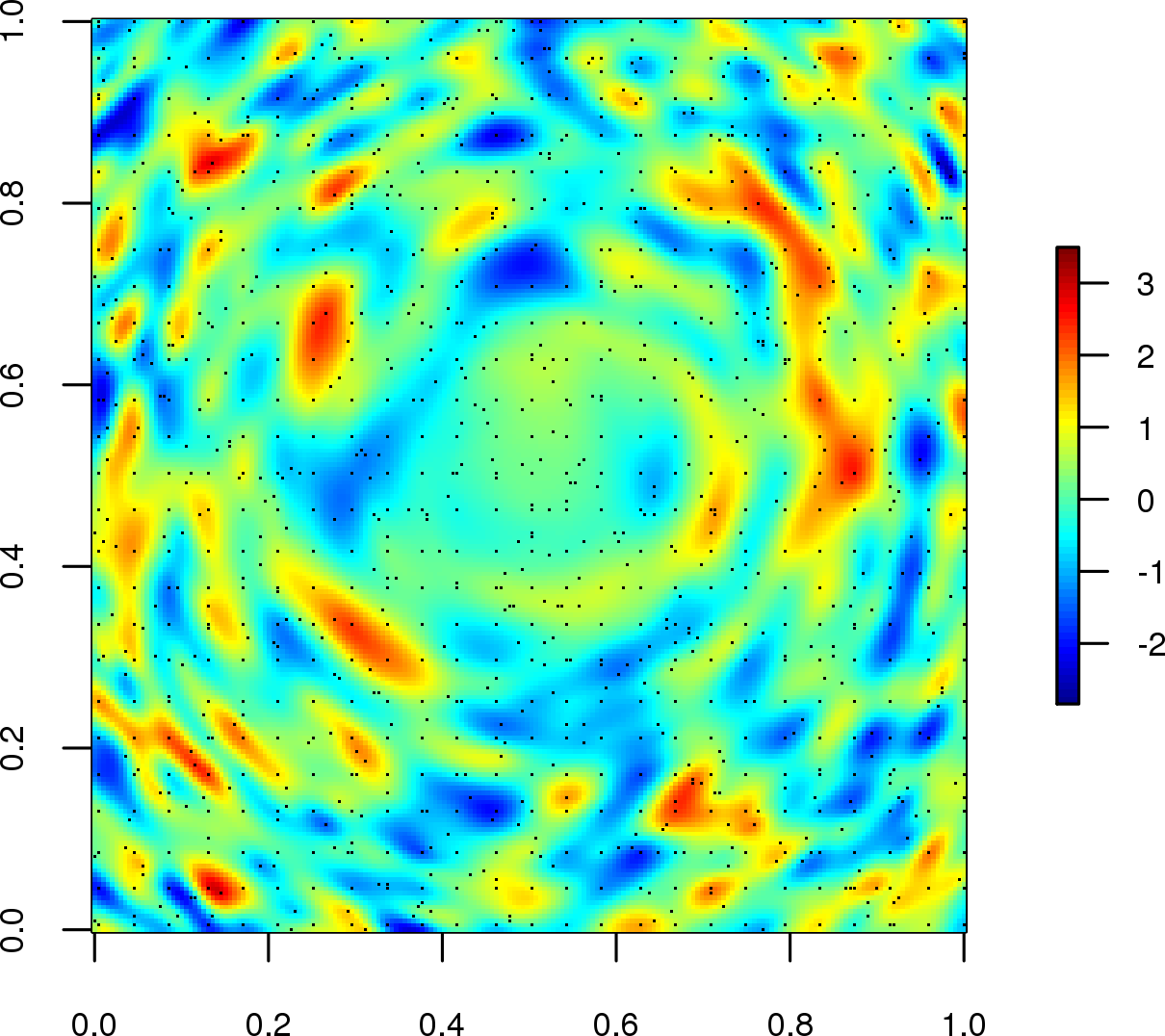}
                \caption{}\label{Fig5c}
        \end{subfigure}
        \medskip
        \begin{subfigure}[h!]{0.28\textwidth}
                \centering
                \includegraphics[width=1\textwidth]{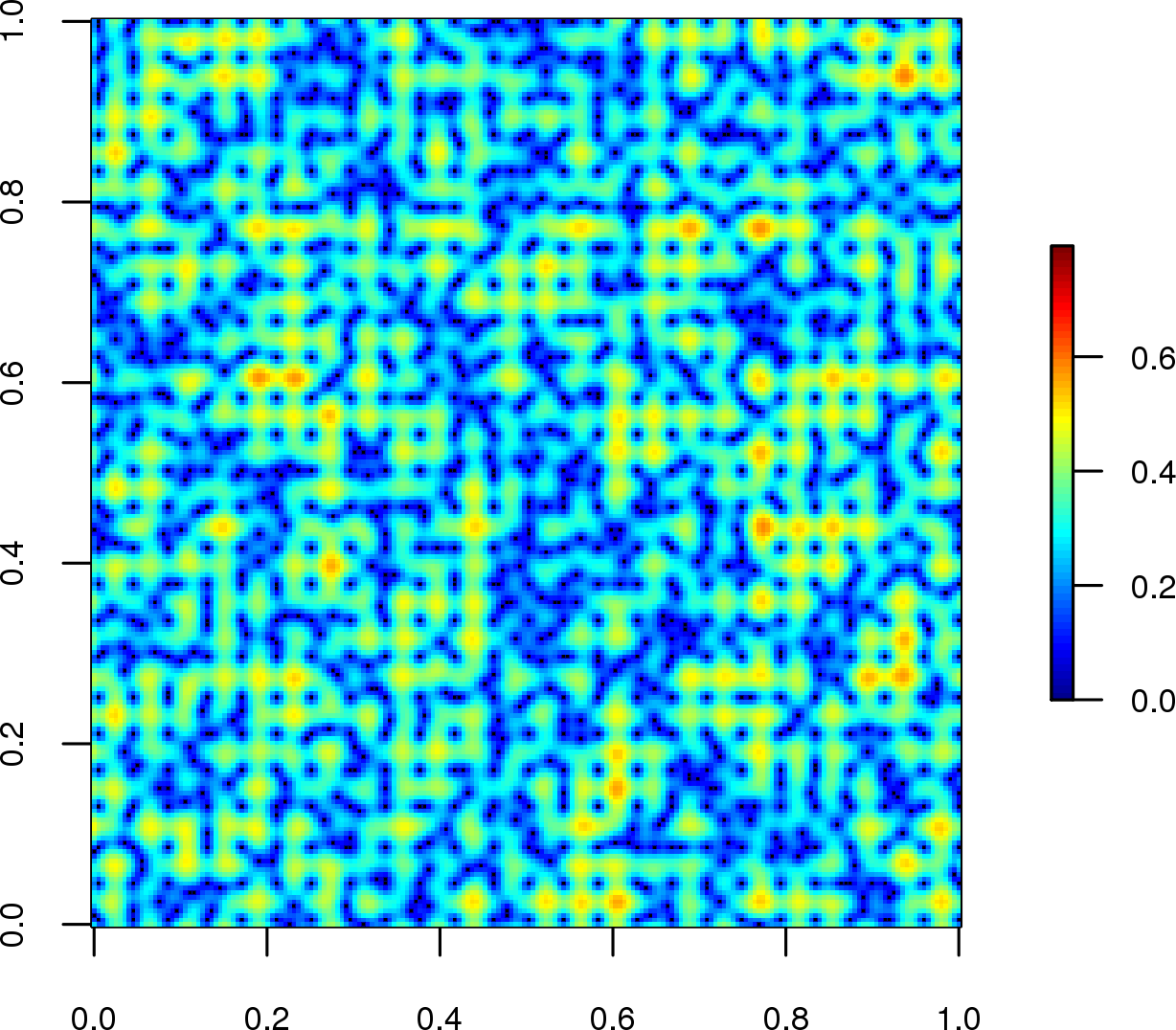}
                \caption{}\label{Fig5d}
        \end{subfigure}
        \quad
        \begin{subfigure}[h!]{0.28\textwidth}
                \centering
                \includegraphics[width=1\textwidth]{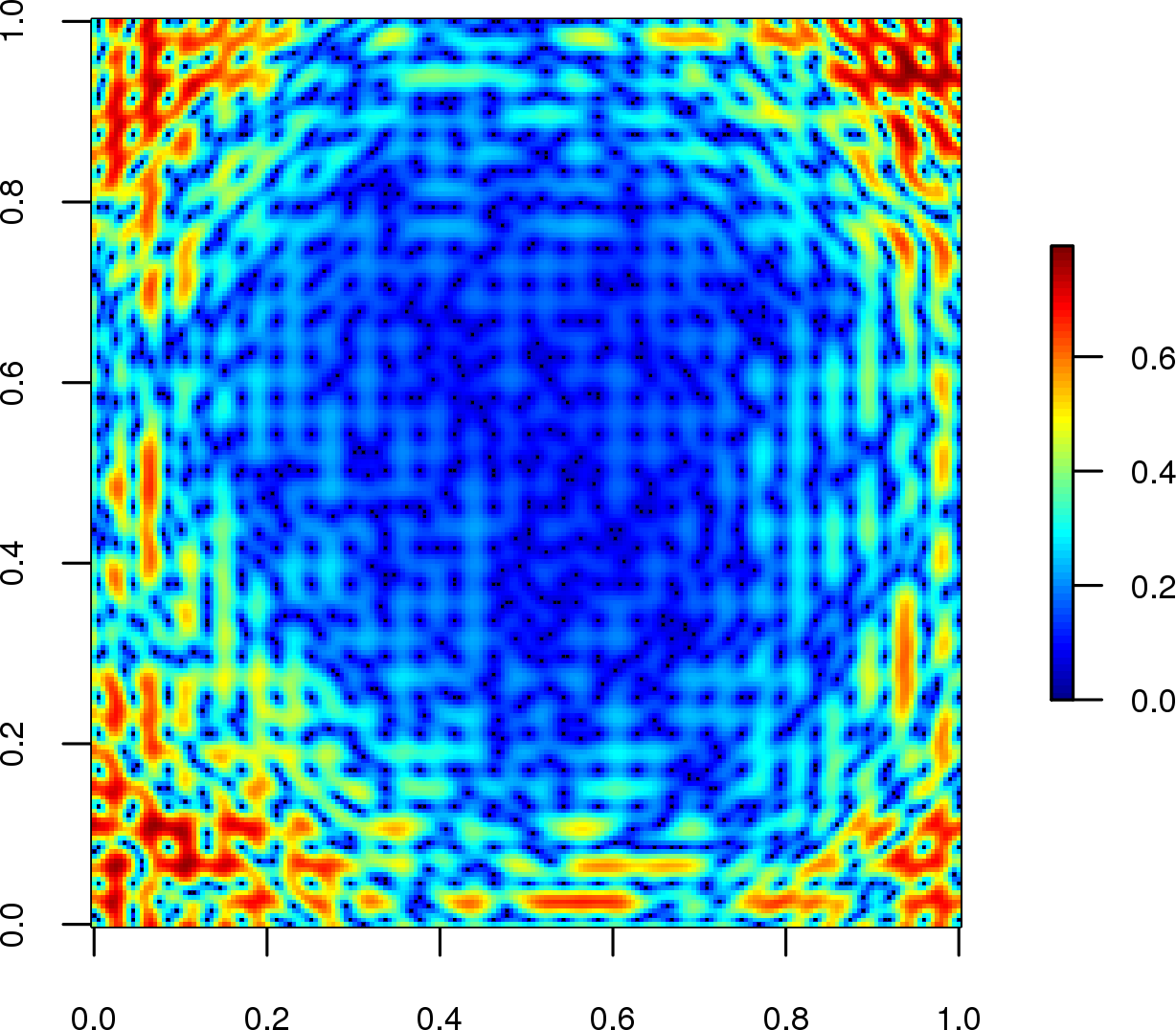}
                \caption{}\label{Fig5e}
        \end{subfigure}
        \quad
        \begin{subfigure}[h!]{0.28\textwidth}
                \centering
                \includegraphics[width=1\textwidth]{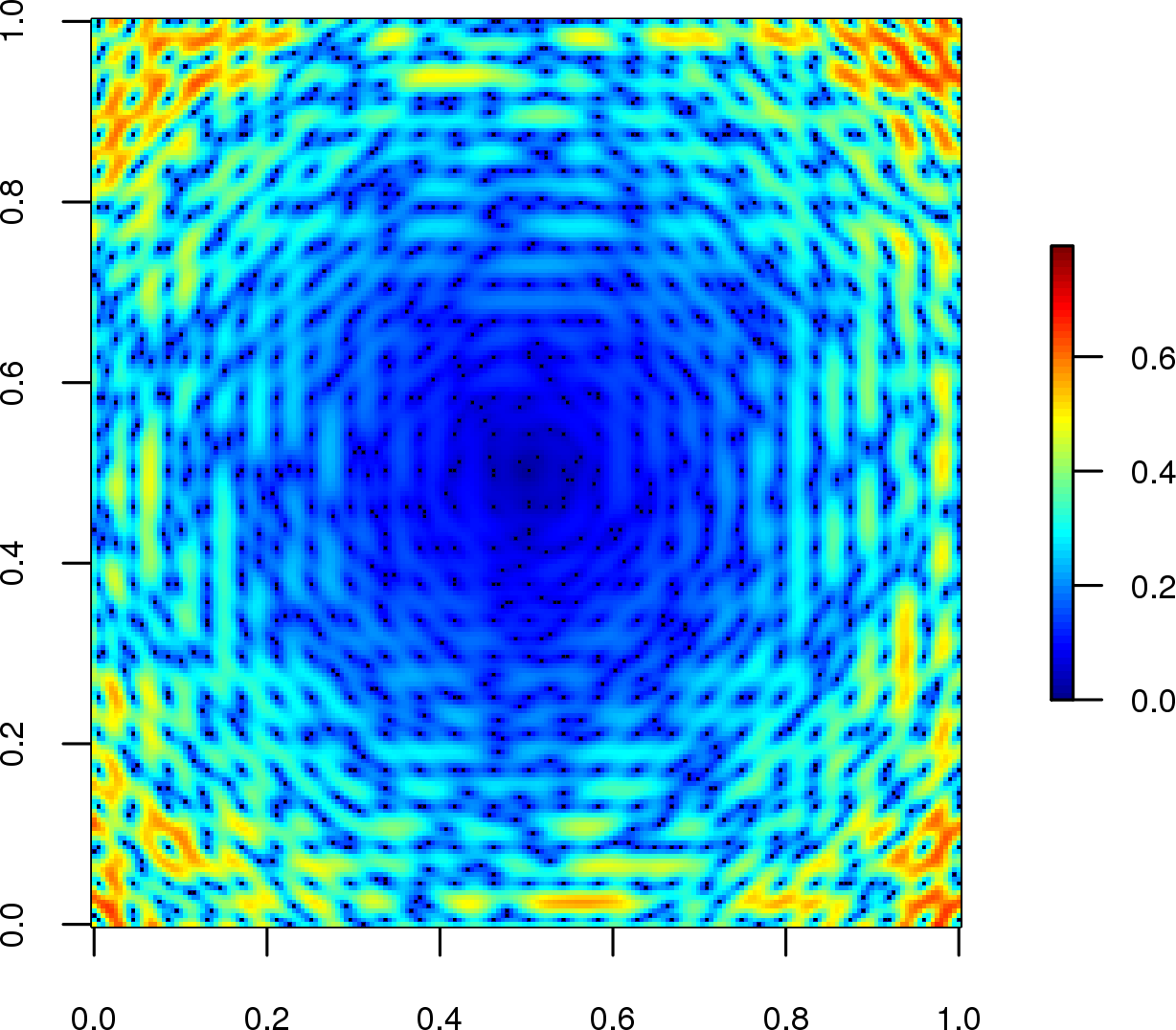}
                \caption{}\label{Fig5f}
        \end{subfigure}
        
        \caption{Predictions for the estimated stationary model, estimated non-stationary model and reference non-stationary model (a, b, c). Prediction standard deviations for the stationary model, estimated non-stationary model and reference non-stationary model (d, e, f).}\label{Fig5}
\end{figure}

The predictions and prediction standard deviations for the estimated stationary, estimated non-stationary and reference models are shown in figure \ref{Fig5}. The differences between kriging standard deviation maps  are visibly marked. The stationary method provides an oversimplified version of the kriging standard deviations because it does not take into account variations of variability.

\section{Application to soil data}
\label{sec6}
We now study a dataset coming from a soil gamma radiometric potassium concentration survey in the region of the Hunter Valley, NSW, Australia. This data set was used by \citet{McB13} to illustrate their approach. We have a  training data (537 points) which serves to calibrate the model and validation data (1000 points) which serves only to assess prediction performance. A configuration of $125$ anchor points are used to determine the deformed space. We compare ordinary kriging schemes with stationary and non-stationary assumptions.

\begin{figure}[H]
        \centering
        \begin{subfigure}[h]{0.28\textwidth}
                \centering
                \includegraphics[width=1\textwidth]{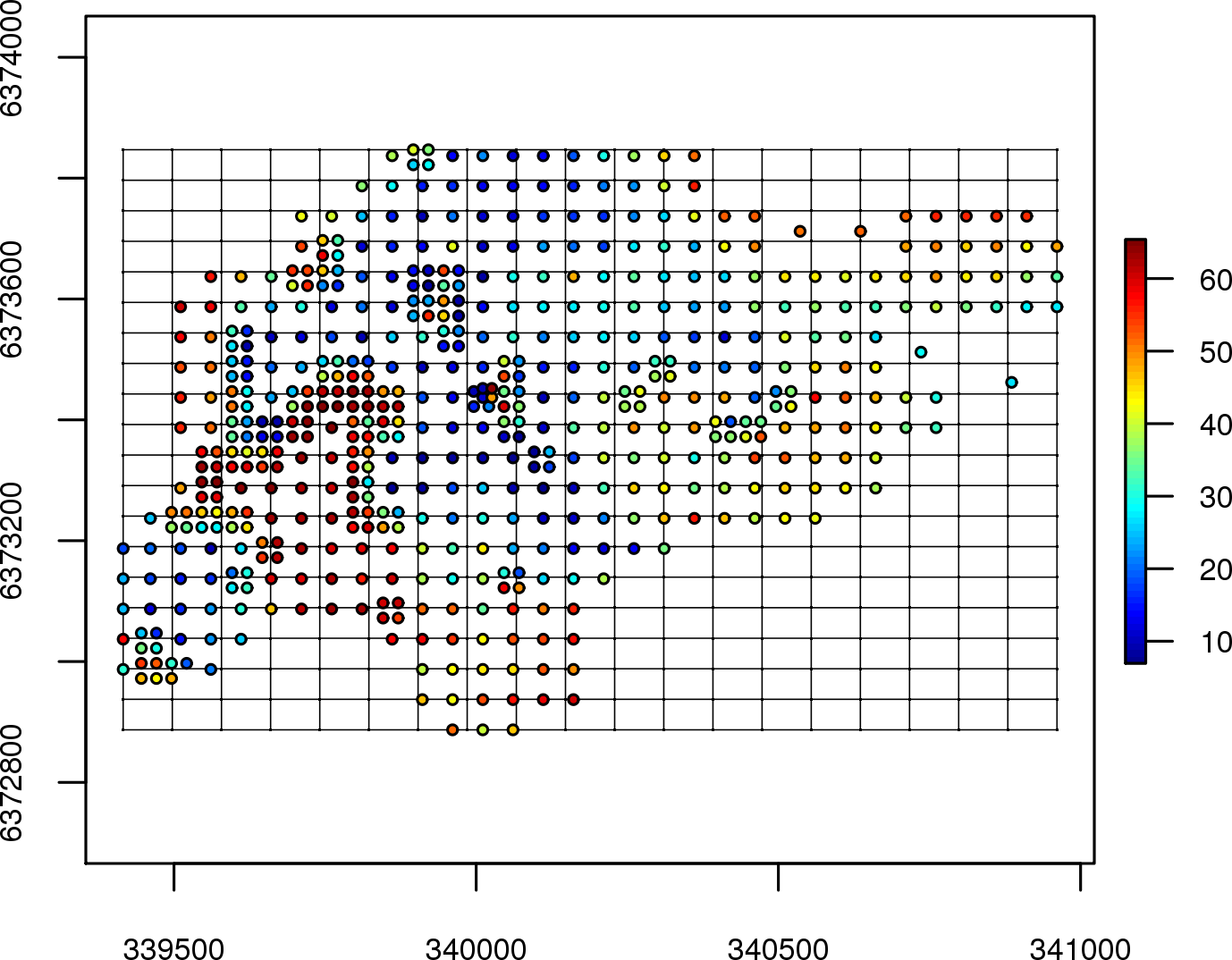}
                \caption{}\label{Fig6a}
        \end{subfigure}
        \qquad  \qquad
        \begin{subfigure}[h]{0.28\textwidth}
                \centering
                \includegraphics[width=1\textwidth]{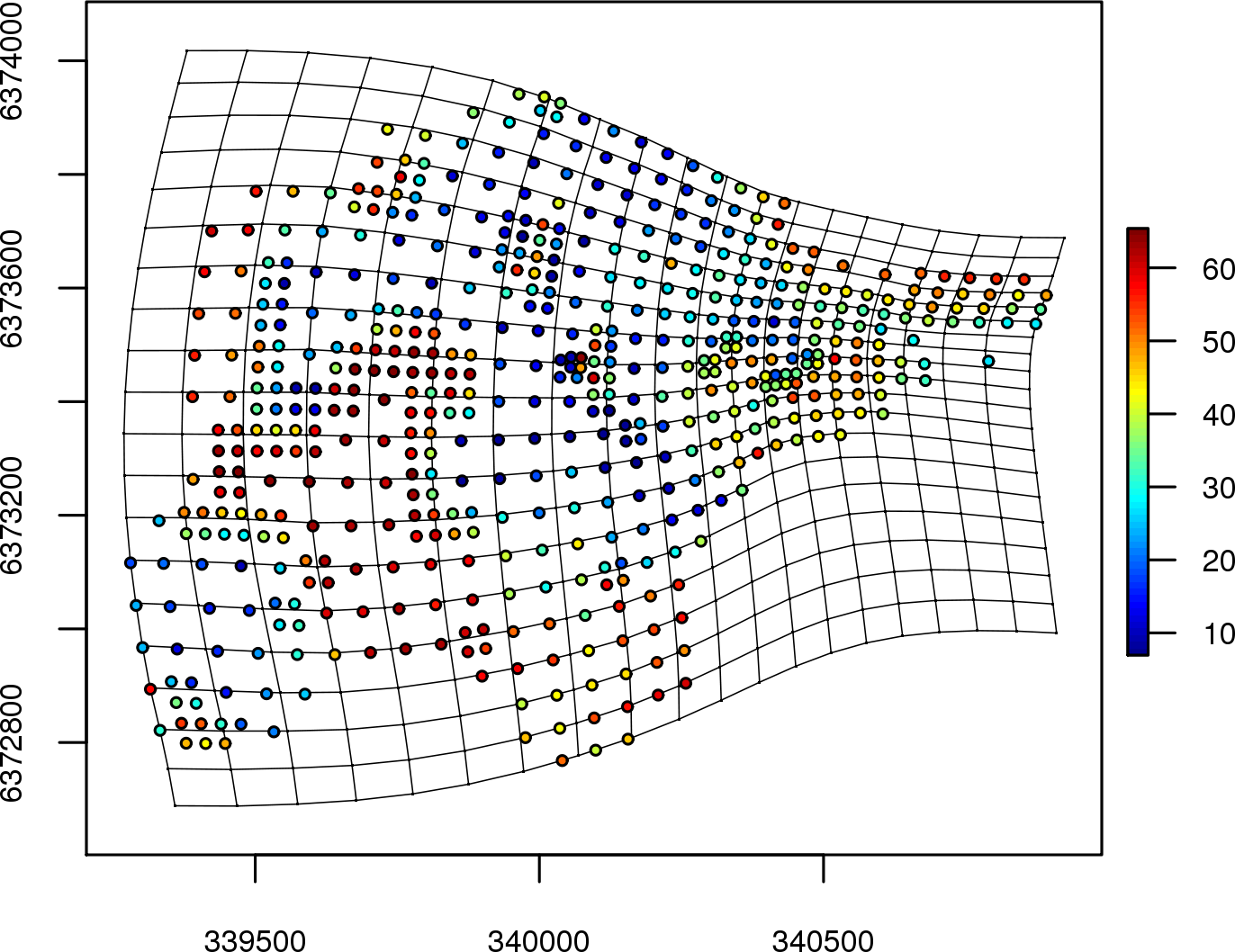}
                \caption{}\label{Fig6b}
        \end{subfigure}
        
        \medskip
                \begin{subfigure}[h]{0.28\textwidth}
                \centering
                \includegraphics[width=1\textwidth]{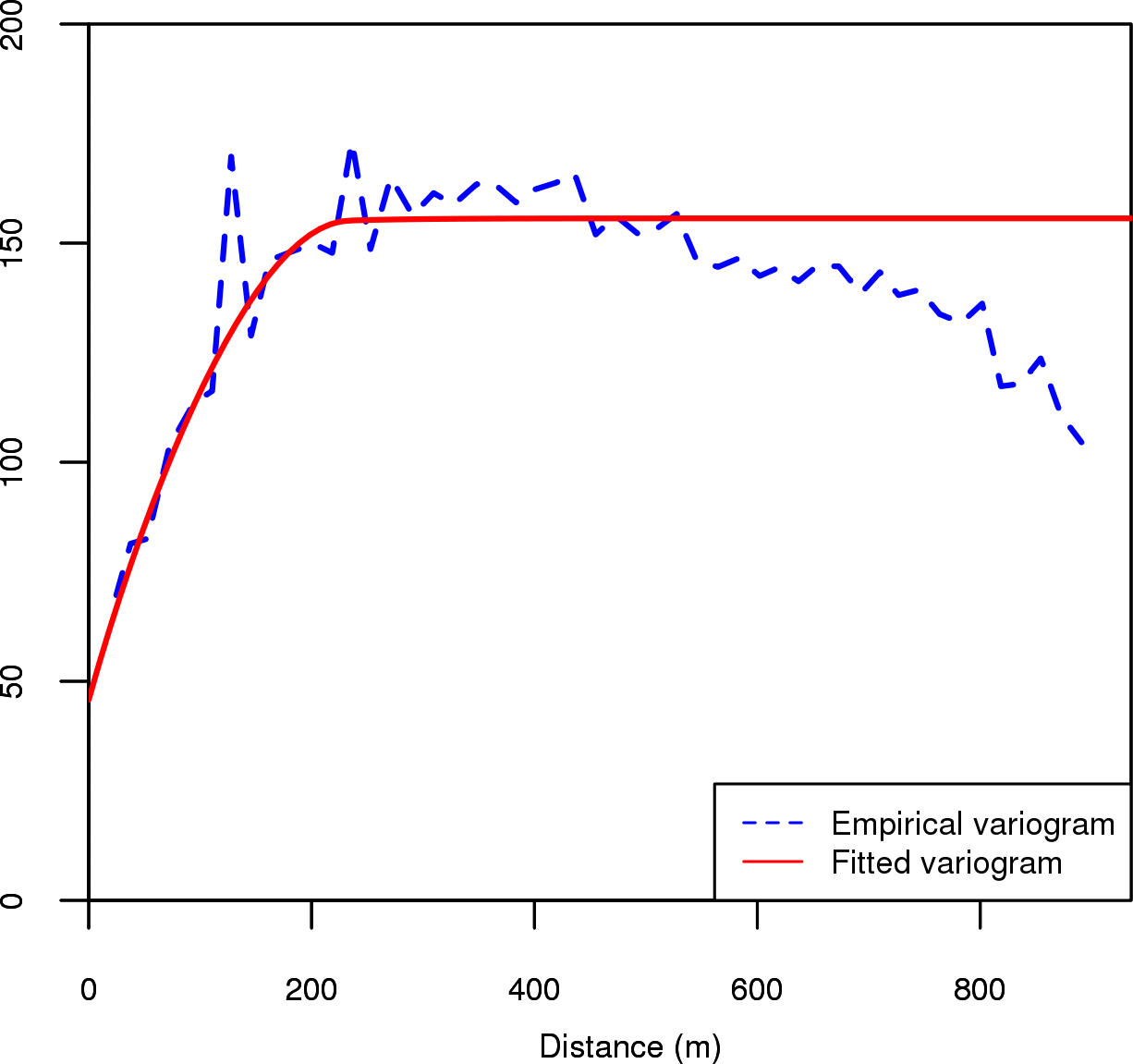}
                \caption{}\label{Fig6c}
        \end{subfigure}
        \qquad  \qquad
        \begin{subfigure}[h]{0.28\textwidth}
                \centering
                \includegraphics[width=1\textwidth]{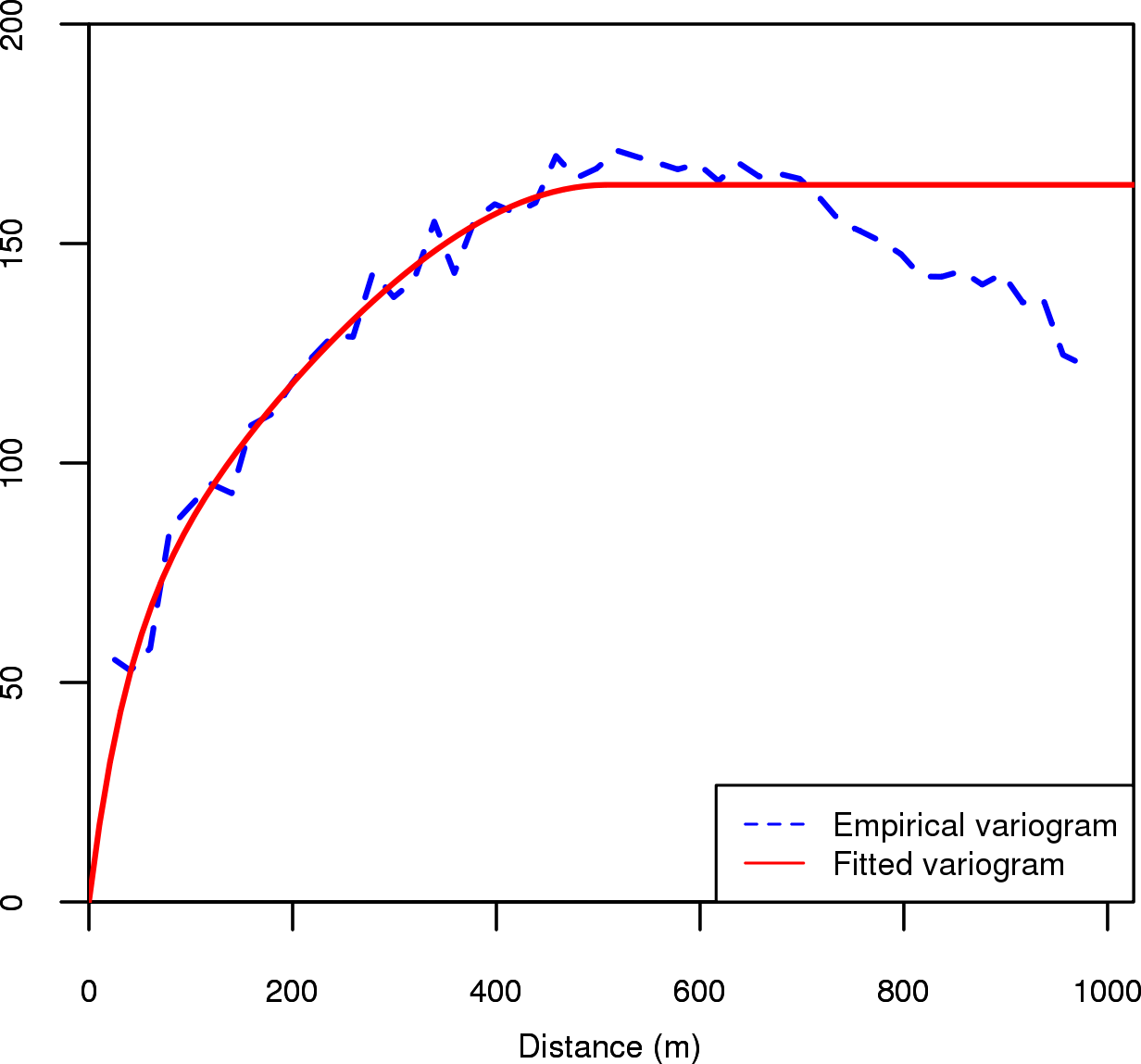}
                \caption{}\label{Fig6d}
        \end{subfigure}
        
        \caption{(a, c) Gamma K concentration data in geographical space, and the estimated stationary variogram model. (b, d) Gamma K concentration data in the deformed space and the estimated transformed variogram model.}\label{Fig6}
\end{figure}

\begin{figure}[H]
        \centering
\begin{subfigure}[h!]{0.28\textwidth}
                \centering
                \includegraphics[width=1\textwidth]{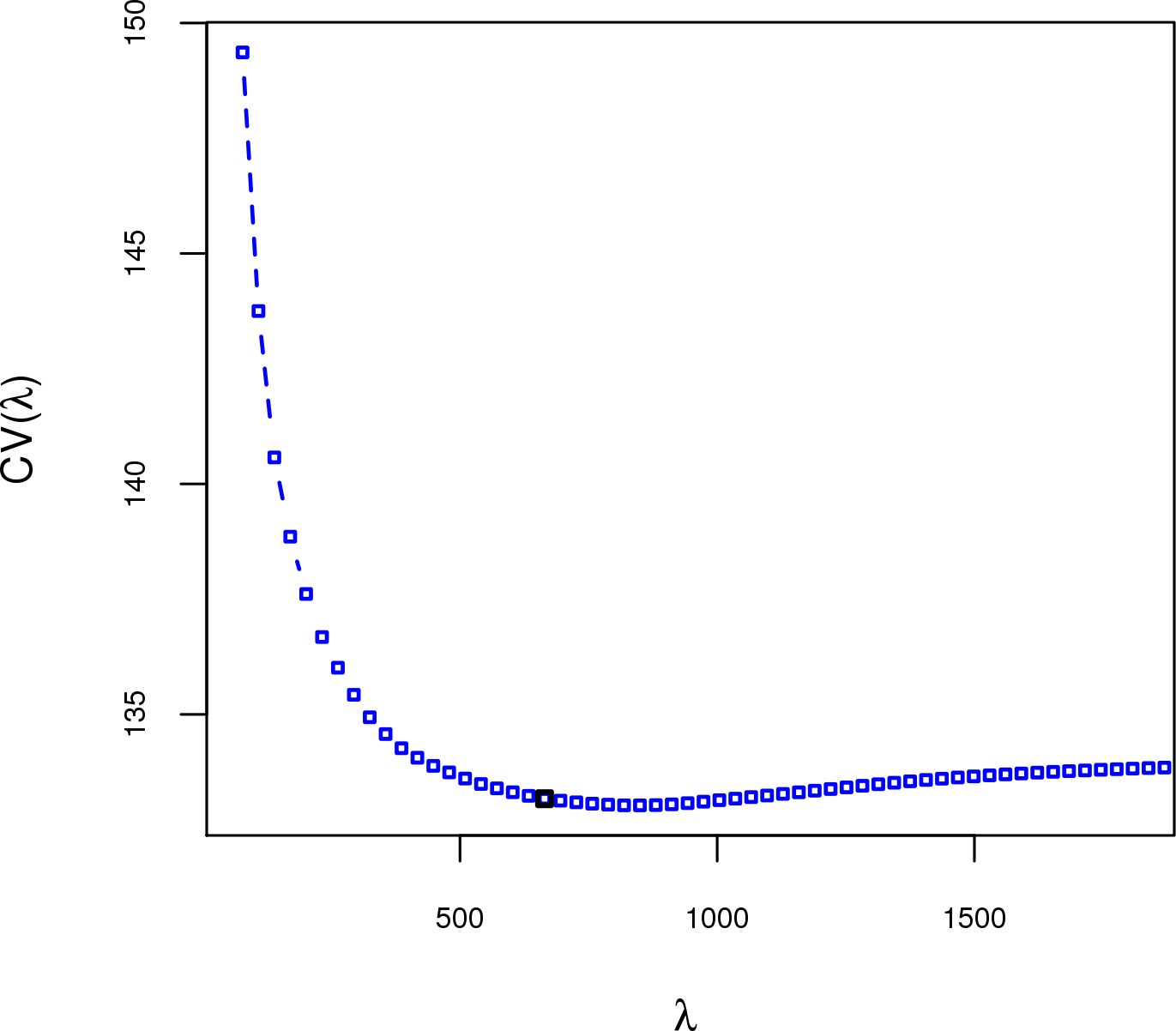}
                \caption{}\label{Fig7a}
        \end{subfigure}
        \qquad
        \begin{subfigure}[h!]{0.30\textwidth}
                \centering
                \includegraphics[width=1\textwidth]{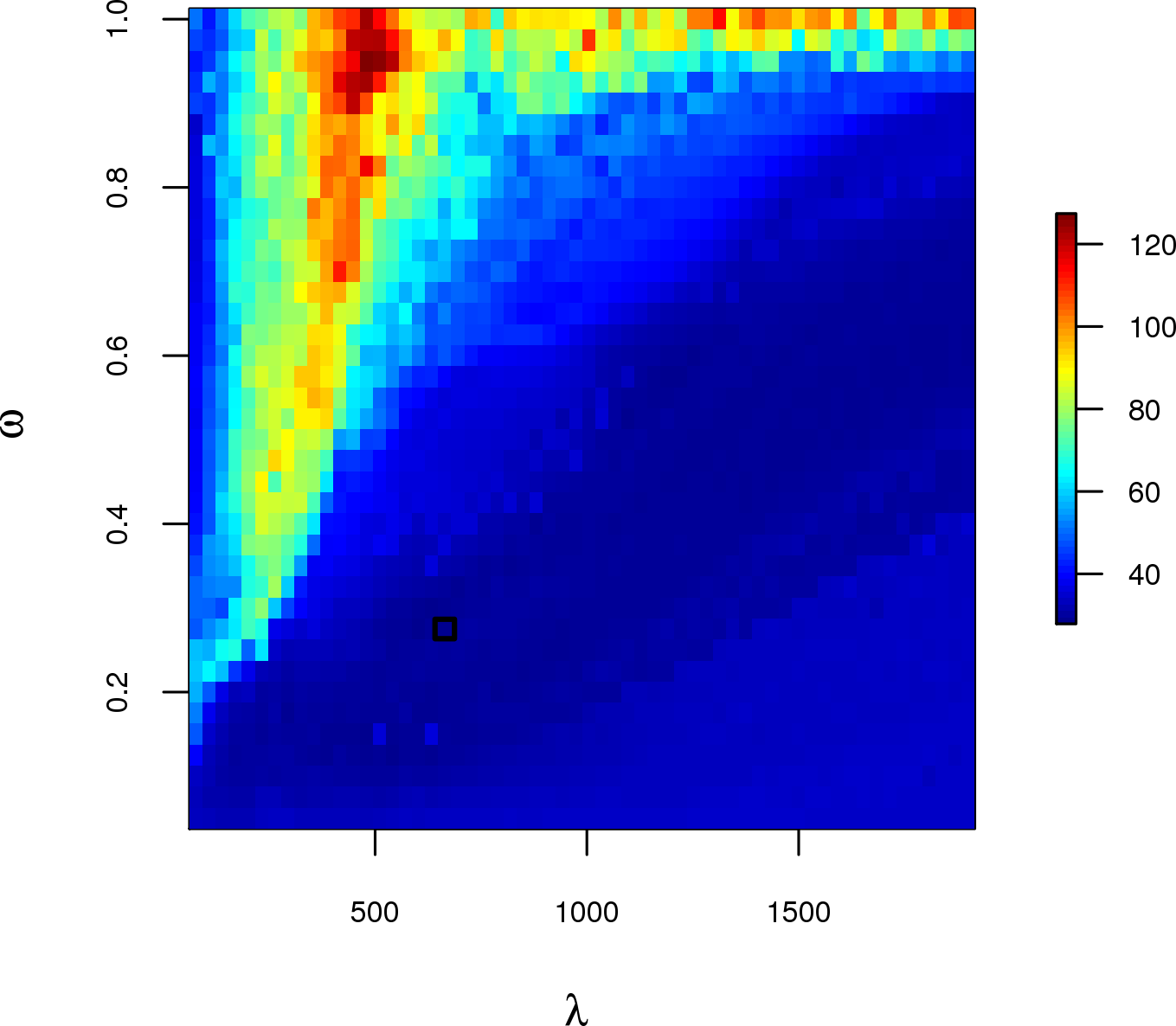}
                \caption{}\label{Fig7b}
        \end{subfigure}
        \caption{(a, b) Cross-validation score functions $CV_1(\lambda)$ and $CV_2(\lambda,\omega)$.}\label{Fig7}
\end{figure}

Figures \ref{Fig6a} and \ref{Fig6b} show respectively the data points in the geographical space and their image in the deformed space. We observe that the deformation shrinks the geographical space in the north-east region while stretchs it in other regions. This means that the north-east region corresponds to a region of relatively high spatial correlation. The cross-validation score functions are presented in figures \ref{Fig7a} and \ref{Fig7b}. The optimal value of hyper-parameters correspond to  $\lambda=664$ m and $\omega=0.275$. Figures \ref{Fig6c} and \ref{Fig6d} respectively  present the variogram correspond to a stationary model in the geographical and deformed spaces. The estimated variogram models shown in figures \ref{Fig6c} and \ref{Fig6d} are respectively:
\begin{eqnarray}\label{Eq14}
\widehat \gamma_{1}(\|\mathbf{h}\|)&=& 46\times\delta_0(\|\mathbf{h}\|) + 21\times\mbox{Exp}_{187}(\|\mathbf{h}\|) + 89\times\mbox{Sph}_{234}(\|\mathbf{h}\|),\\
\widehat \gamma_{0}(\|\mathbf{h}\|)&=& 62\times\mbox{Exp}_{101}(\|\mathbf{h}\|) + 102\times\mbox{Sph}_{428}(\|\mathbf{h}\|).
\end{eqnarray}

\begin{figure}[H]
        \centering
        \begin{subfigure}[h!]{0.28\textwidth}
                \centering
                \includegraphics[width=1\textwidth]{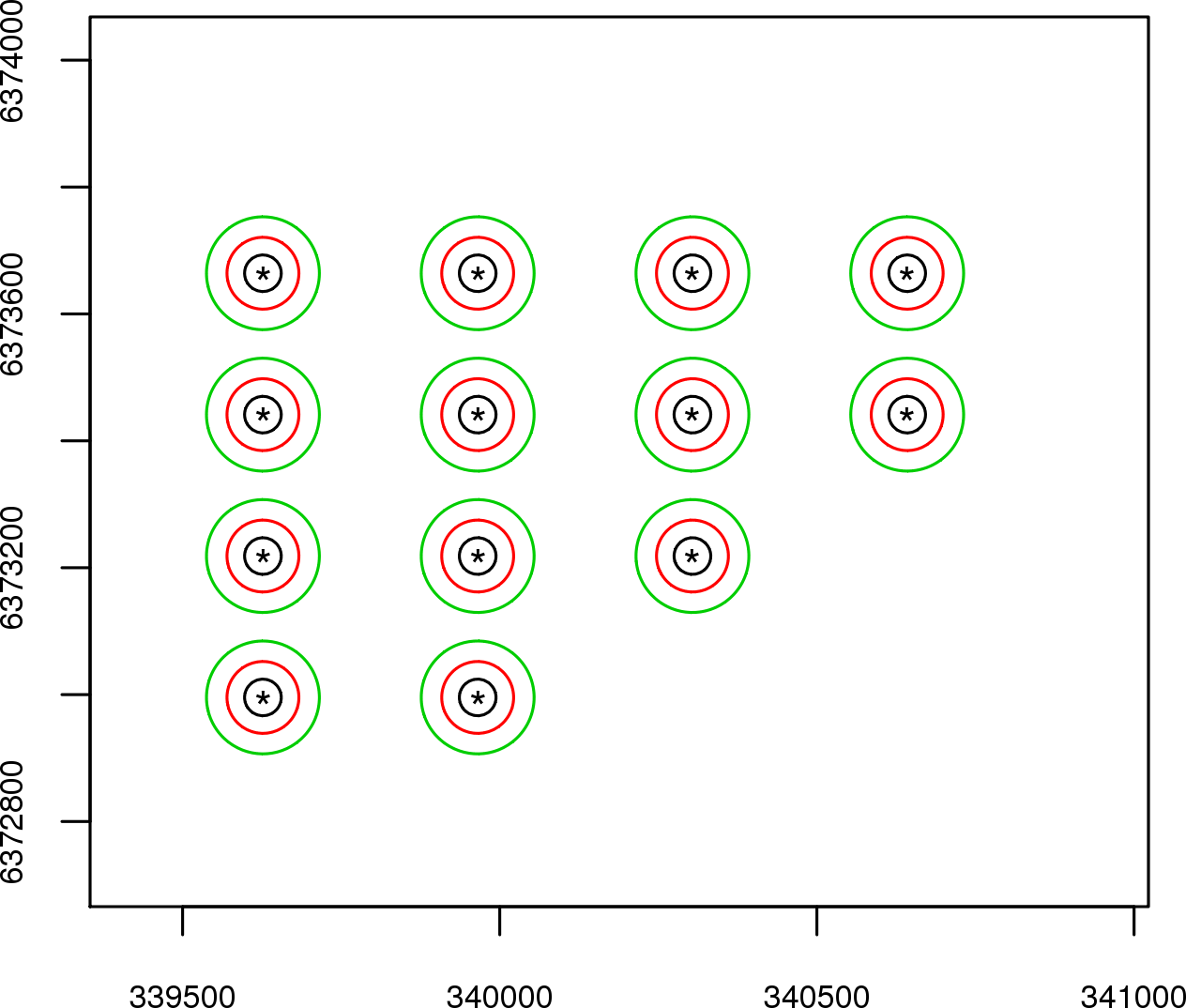}
                \caption{}\label{Fig8a}
        \end{subfigure}
        \quad
        \begin{subfigure}[h!]{0.28\textwidth}
                \centering
                {\includegraphics[width=1\textwidth]{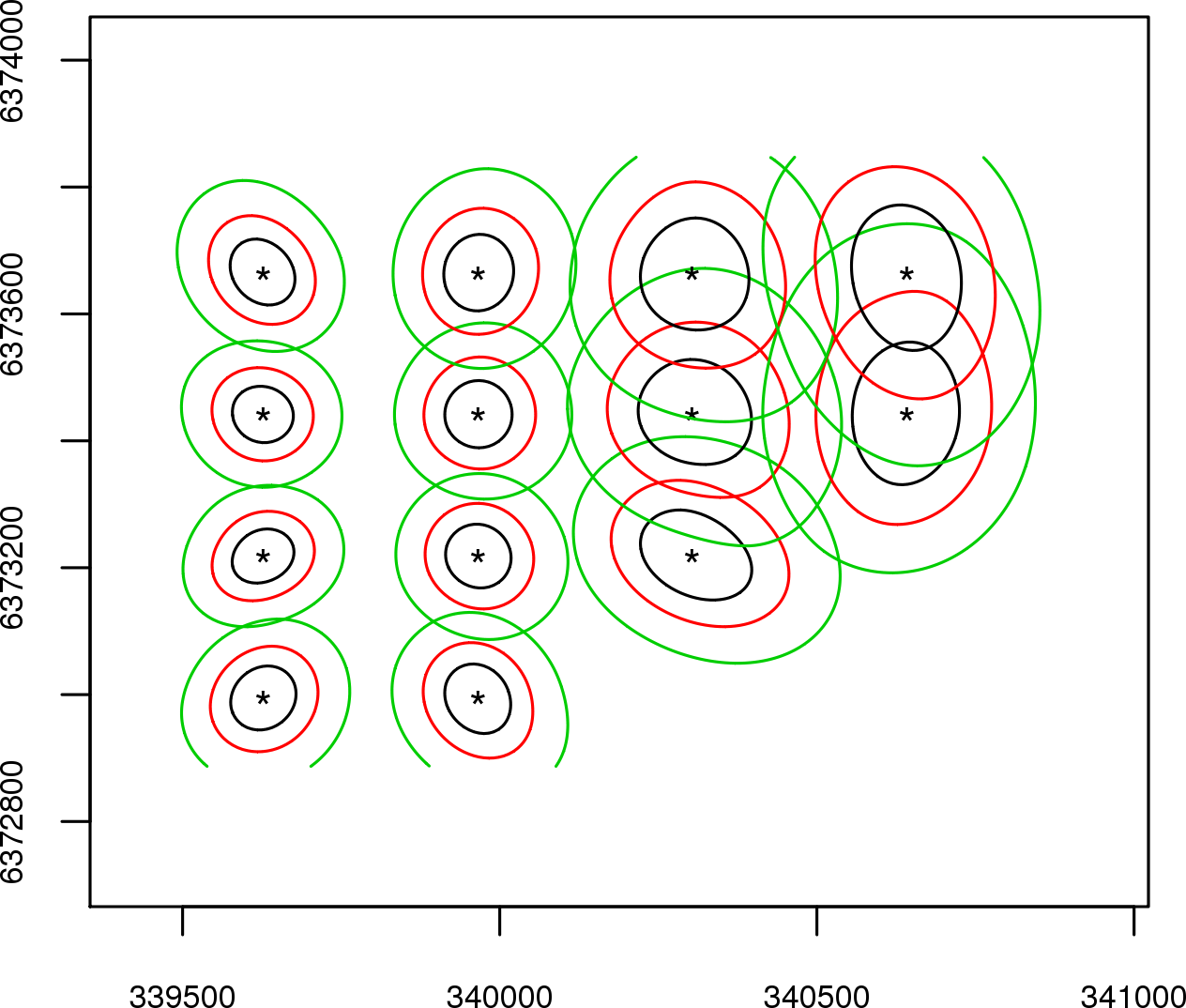}}
                \caption{}\label{Fig8b}
        \end{subfigure}
               
        \caption{ Variogram  level contours at few points for the estimated stationary and non-stationary models (a, b). Level contours correspond to the values: $70$ (black), $90$ (red) and $110$ (green).}\label{Fig8}
\end{figure}

A visualization of the variogram at certain points for estimated stationary and non-stationary models is presented in figure \ref{Fig8}. We can see how the non-stationary spatial dependence structure changes the shape from one place to another as compared to the stationary one.

\begin{figure}[H]
        \centering
        \begin{subfigure}[h!]{0.30\textwidth}
                \centering
                \includegraphics[width=1\textwidth]{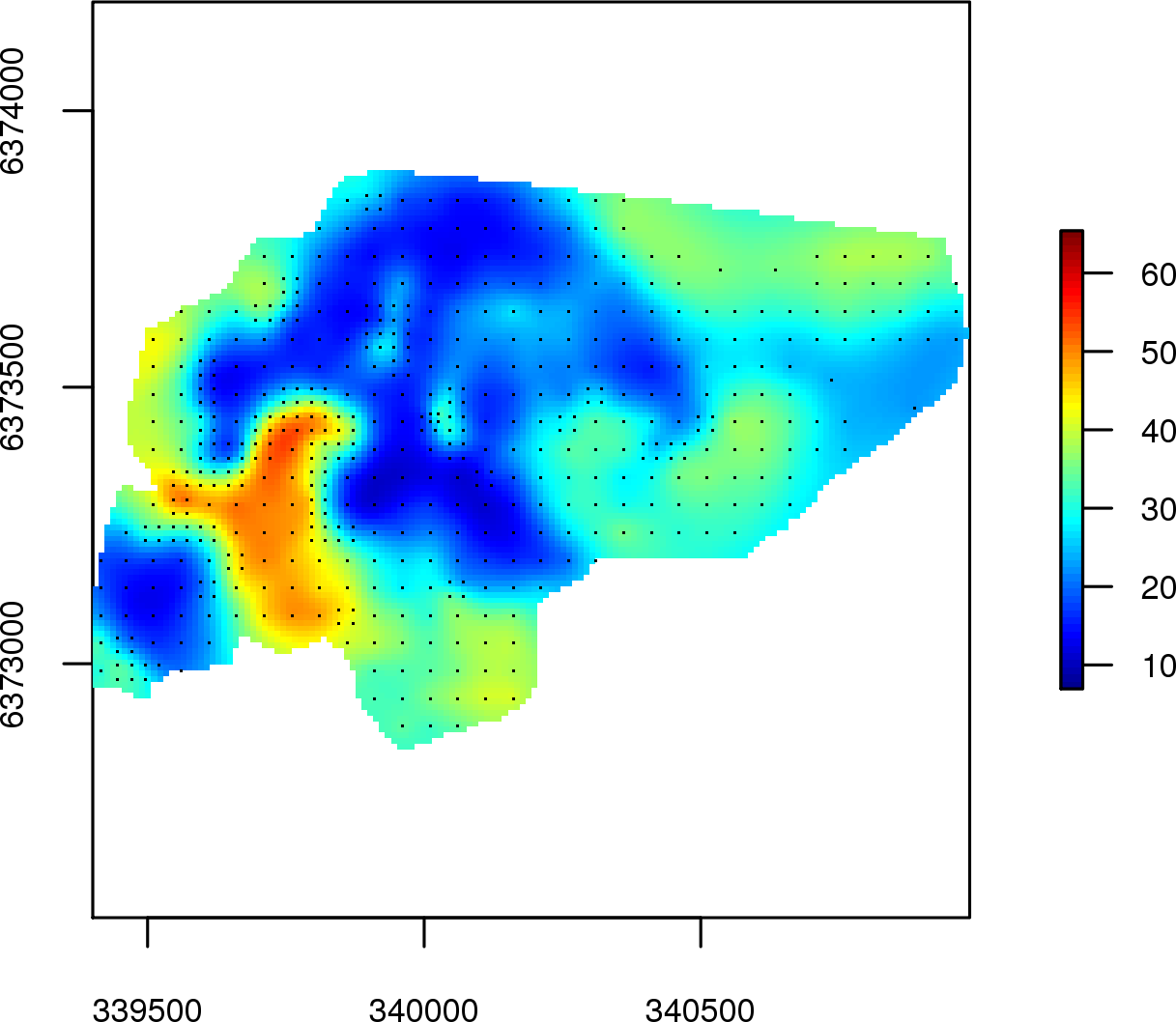}
                \caption{}\label{Fig9a}
        \end{subfigure}
        \quad 
        \begin{subfigure}[h!]{0.30\textwidth}
                \centering
                \includegraphics[width=1\textwidth]{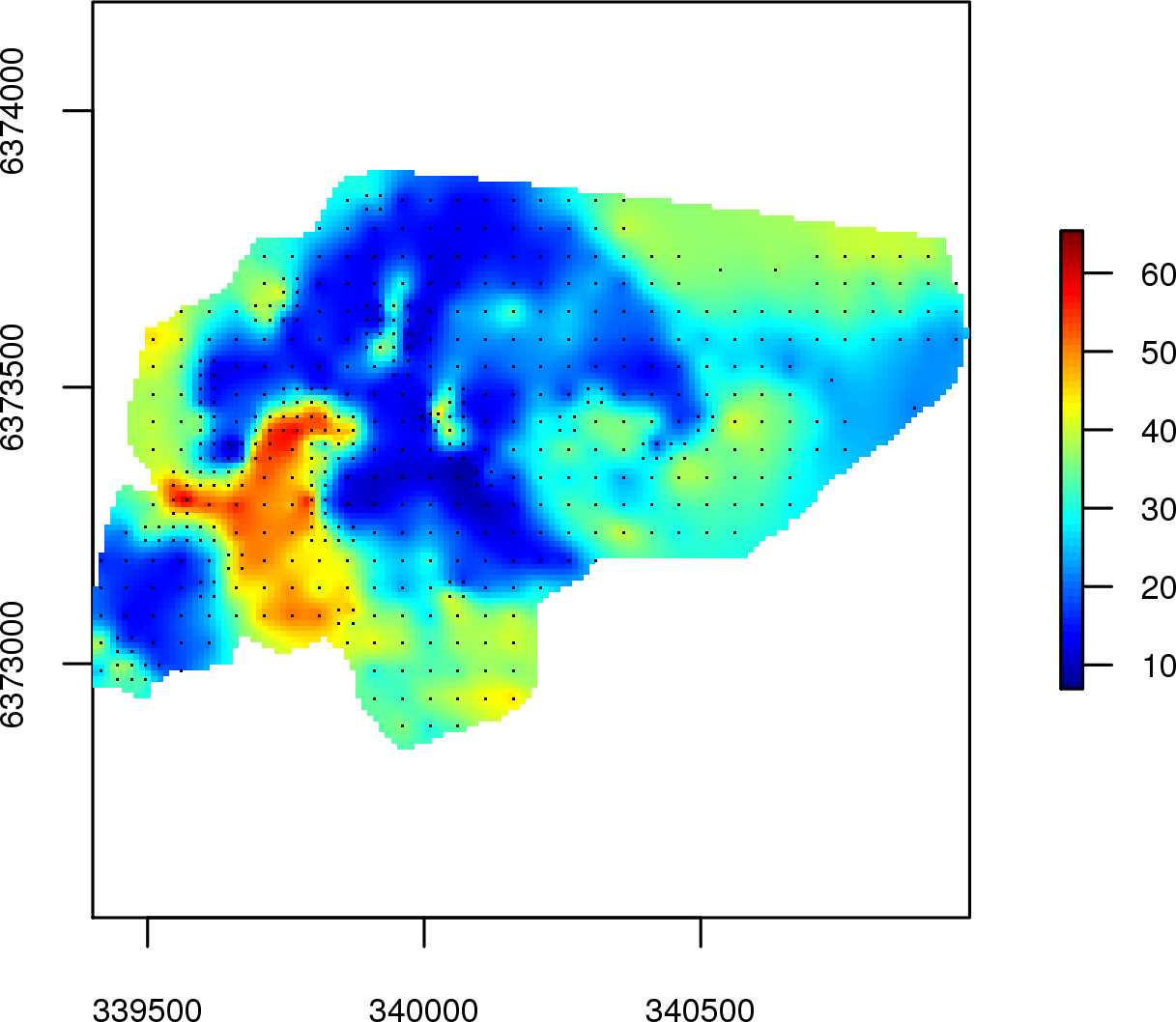}
                \caption{}\label{Fig9b}
        \end{subfigure}
        
        \medskip
        
        \begin{subfigure}[h!]{0.30\textwidth}
                \centering
                \includegraphics[width=1\textwidth]{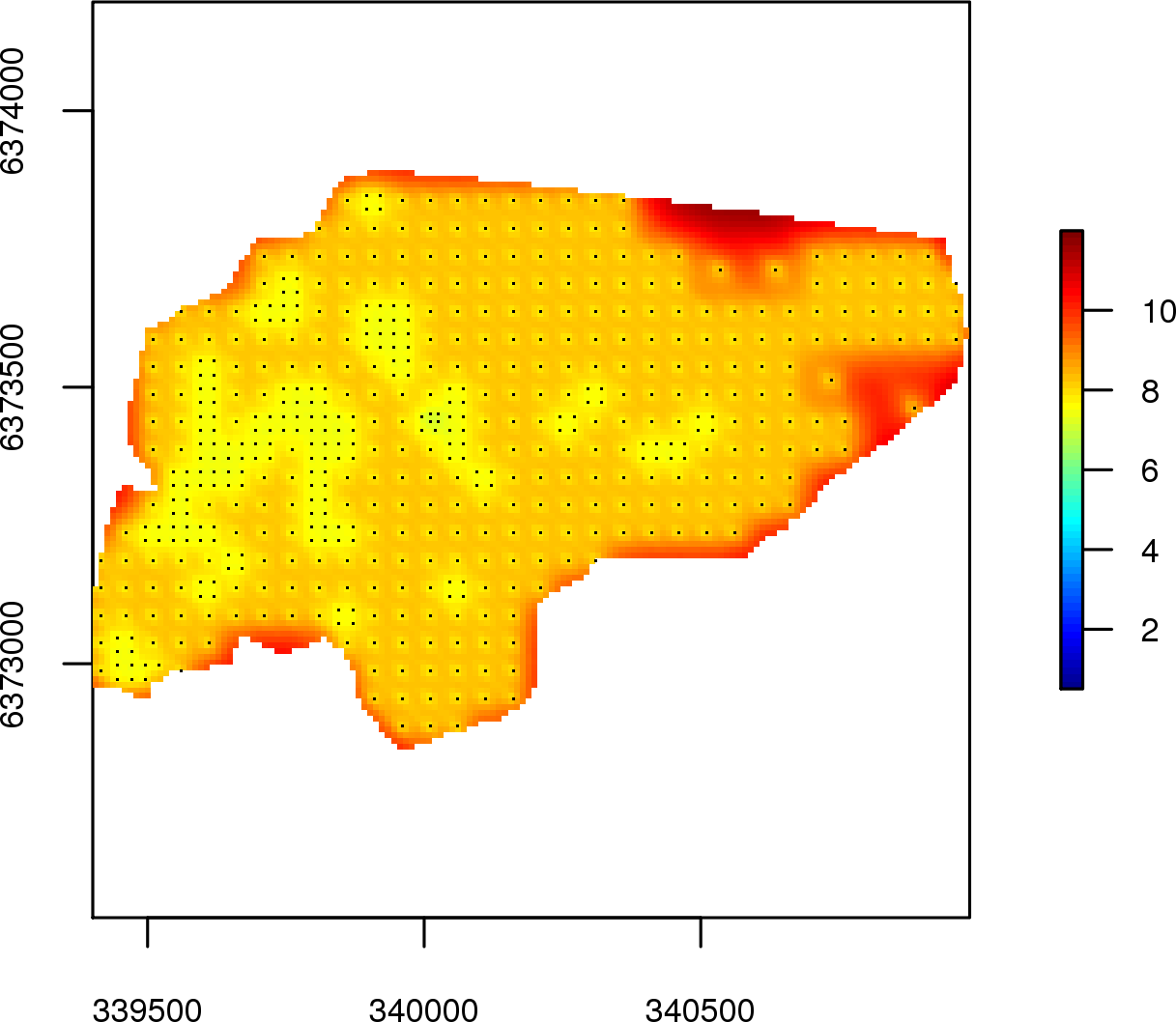}
                \caption{}\label{Fig9c}
        \end{subfigure}
        \quad 
        \begin{subfigure}[h!]{0.30\textwidth}
                \centering
                \includegraphics[width=1\textwidth]{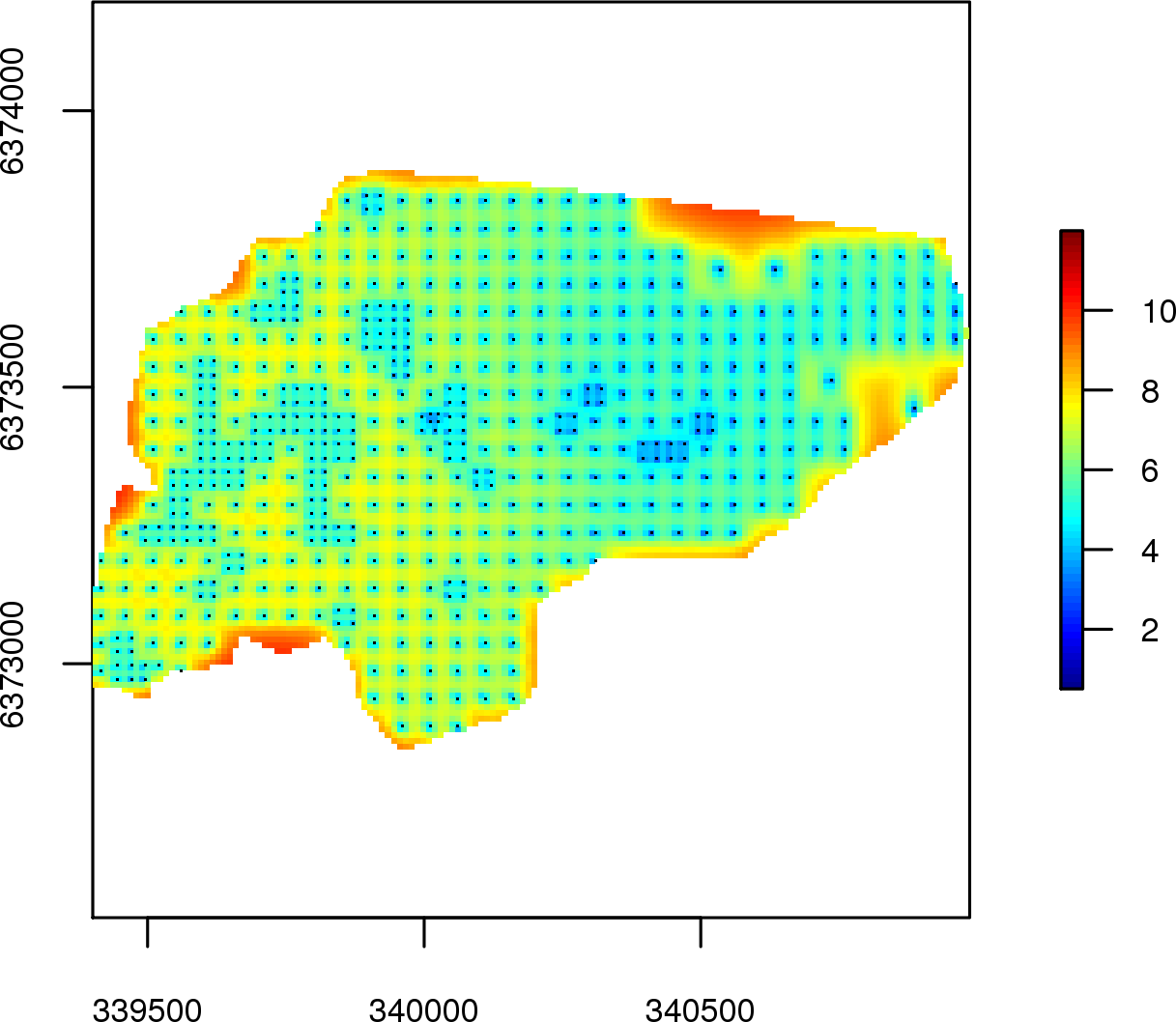}
                \caption{}\label{Fig9d}
        \end{subfigure}
        \caption{(a, b) Predictions and prediction standard deviations for the estimated stationary model for gamma K concentration. (c, d) Predictions and prediction standard deviations for the estimated  non-stationary model for gamma K concentration.}\label{Fig9}
\end{figure}

The performance statistics computed on the validation data set presented in table \ref{Tab2} show that the proposed approach outperforms the stationary one. The cost of non-using the non-stationary approach in this case is not negligible: in average the prediction  at validation locations is about 21\% better for the estimated non-stationary model than for the estimated stationary model, in terms of RMSE. The kriged values and kriging standard deviations for the estimated stationary and non-stationary models are shown in figure \ref{Fig9}. The overall look of the predicted values and prediction standard deviations associated with each model differ. In particular, the proposed method takes into account certain local characteristics of the regionalization that the stationary approach is unable to retrieve.

\begin{table}[H]
\begin{center}
\begin{tabular}{lcc}
\hline
         & Stationary Model& Non-Stationary Model  \\
         \hline
    
    Mean Absolute Error   & 2.86   & 2.17 \\
              
    Root Mean Square Error  & 3.88    & 3.21 \\
    
    Normalized Mean Square Error &0.19  & 0.26   \\
    
    Logarithmic Score & 6400   & 5516   \\
    
    Continued Rank Probability Score & 4.97   & 3.61  \\
    \hline
\end{tabular}
\end{center}
\caption{Predictive performance statistics on a validation set of 1000  locations.}
\label{Tab2}
\end{table}

\section{Conclusions and future work}
\label{sec7}
The proposed approach appears to be an efficient tool to estimate a non-stationary spatial dependence structure from a single realization. It improves the estimation of a Random Function with non-stationary spatial dependence structure, as illustrated by the simulated examples and the real application. Additionally, the non-stationary approach leads to better prediction than the stationary model as measured by predictive scores. 

One advantage of the proposed approach is that instead of finding the deformation using all data points, we can do it with a reduced number of points, compared to existing methods. This is a major improvement for large datasets. Indeed, the use of anchor points makes it possible to run the NMDS step even for very large datasets. Moreover, the approach incorporates spatial constraints which guarantee the bijection property  of the deformation. This point was actually a major drawback while trying to fit a deformation function. 

The proposed approach is easy to implement and allows to use all developments already made in the stationary framework through the deformation. Indeed, prediction and simulation under stationarity is a well understood subject, for which fast and robust techniques are available. The approach also provides an exploratory analysis tool for the non-stationarity. In fact, the spatial deformation encodes the non-stationarity and the representation of the deformed space allows to identify the regions of strong and weak continuity.

One future direction of research would be to investigate the statistical properties (such as consistency) of the non-parametric kernel estimator of the non-stationary variogram. This would have to be done in an asymptotic context by following the work of \citet{Gar04} in the stationary framework. 

The deformation mapping need not to be restricted to the class of functions that are currently use. From our experience, thin-plate spline radial basis functions work well but a different interpolation method may prove useful. It would be also interesting to account for covariate information in the space deformation model, in order to improve the estimation of the deformation. This would have to be done by following the work of \citet{Sch11}. 

One challenge remaining is the selection of hyper-parameters. Indeed, the cross-validation procedures presented here remains computationally demanding. 

\section*{Acknowledgements}
The authors would like to thank Dr Budiman Minasny at the Faculty of Agriculture \& Environment at the University of Sydney in Australia, for providing the data used in this paper.

\appendix
\section{}
\label{appendix1}

\textit{Proof of proposition \ref{Prop1}}

Consider a set of points $\{\mathbf{x}_1,\ldots,\mathbf{x}_n\} \subset G$ and a set of real numbers $\{\lambda_1,\ldots,\lambda_n\}$ such that $\sum_{i=1}^n\lambda_i=0, n \in \mathds{N}^*$. Let $\mathbf{x}_i^*=f(\mathbf{x}_i), \forall i=1,\ldots,n$. We have
\begin{eqnarray*}
\sum_{i=1}^n\sum_{j=1}^n\lambda_i\lambda_j\gamma_{0} \circ (f \times f)(\mathbf{x}_i,\mathbf{x}_j)
&=& \sum_{i=1}^n\sum_{j=1}^n\lambda_i\lambda_j\gamma_{0}(f(\mathbf{x}_i),f(\mathbf{x}_j))\\
&=&\sum_{i=1}^n\sum_{j=1}^n\lambda_i\lambda_j\gamma_{0}(\mathbf{x}_i^*,\mathbf{x}_j^*), \  \mbox{with} \  \mathbf{x}_1^*,\ldots,\mathbf{x}_n^* \in D \  \mbox{and} \sum_{i=1}^n\lambda_i=0\\
&\leq& 0, \ \mbox{since} \  \gamma_0(.) \ \mbox{is a valid variogram on} \  D.
\end{eqnarray*}
Hence $\gamma_{0} \circ (f \times f)$ is a valid variogram on $G$.

\textit{Proof of proposition \ref{Prop2}}

Consider $(\gamma_0,f)$ a solution to \eqref{Eq3}. Let $(\tilde{\gamma_0},\tilde{f})$ such that $\tilde{f}(\mathbf{x})=\mathbf{A}f(\mathbf{x})+\mathbf{b}$ and $\tilde{\gamma}_0(\|\mathbf{u}\|)=\gamma_0(\| \mathbf{A}^{-1}\mathbf{u}\|)$. We have
\begin{eqnarray*}
\tilde{\gamma}_0(\|\tilde{f}(\mathbf{x})-\tilde{f}(\mathbf{y})\|)&=&\tilde{\gamma}_0(\|\mathbf{A}f(\mathbf{x})+\mathbf{b}-\mathbf{A}f(\mathbf{y})-\mathbf{b}\|)\\
&=& \tilde{\gamma}_0(\|\mathbf{A}(f(\mathbf{x})-f(\mathbf{y}))\|)\\
&=& \gamma_0(\|\mathbf{A}^{-1}\mathbf{A}(f(\mathbf{x})-f(\mathbf{y}))\|)\\
&=& \gamma_0(\|f(\mathbf{x})-f(\mathbf{y})\|)
\end{eqnarray*}
Hence $(\tilde{\gamma_0},\tilde{f})$ is a solution as well.

\section*{References}
\bibliographystyle{model1-num-names}

\end{document}